\documentclass[a4paper, 10pt]{article}
\usepackage{amsmath}
\usepackage{amsfonts}
\usepackage{amssymb}
\usepackage{algorithm}
\usepackage{algorithmicx}
\usepackage{algpseudocode}
\usepackage{array}
\usepackage{fixltx2e}
\usepackage{url}
\usepackage{multirow}
\usepackage{float}
\usepackage{lscape}
\usepackage[caption = false]{subfig}
\usepackage{balance}
\usepackage{lscape}
\usepackage{graphicx}
\usepackage{mathrsfs}
\usepackage{bbm}
\usepackage{amsthm}

\newtheorem{theorem}{Theorem}

\begin{document}

%\date{\displaydate{\thedate}}
\title{Avoiding Backtesting Overfitting by Covariance-Penalties: an empirical investigation of the ordinary and total least squares cases}

\author{Adriano~Koshiyama and Nick~Firoozye \\ 
	Department of Computer Science, \\ University College London \\ Gower Street, London WC1E 6BT, United Kingdom \\ 
	\url{[adriano.koshiyama.15, n.firoozye]@ucl.ac.uk}}% <-this % stops a space 

% note the % following the last \IEEEmembership and also \thanks - 
% these prevent an unwanted space from occurring between the last author name
% and the end of the author line. i.e., if you had this:
% 
% \author{....lastname \thanks{...} \thanks{...} }
%                     ^------------^------------^----Do not want these spaces!
%
% a space would be appended to the last name and could cause every name on that
% line to be shifted left slightly. This is one of those "LaTeX things". For
% instance, "\textbf{A} \textbf{B}" will typeset as "A B" not "AB". To get
% "AB" then you have to do: "\textbf{A}\textbf{B}"
% \thanks is no different in this regard, so shield the last } of each \thanks
% that ends a line with a % and do not let a space in before the next \thanks.
% Spaces after \IEEEmembership other than the last one are OK (and needed) as
% you are supposed to have spaces between the names. For what it is worth,
% this is a minor point as most people would not even notice if the said evil
% space somehow managed to creep in.

% make the title area
\maketitle

% As a general rule, do not put math, special symbols or citations
% in the abstract or keywords.
\begin{abstract}
Systematic trading strategies are rule-based procedures which choose portfolios and  allocate assets. In order to attain certain desired return profiles, quantitative strategists must determine a large array of trading parameters. Backtesting, the attempt to identify the appropriate parameters using historical data available, has been highly criticized due to the abundance of misleading results. Hence, there is an increasing interest in devising procedures for the assessment and comparison of strategies, that is, devising schemes for preventing what is known as  {\em backtesting overfitting.} So far, many financial researchers have proposed different ways  to tackle this problem that can be broadly categorised in three types: Data Snooping, Overestimated Performance, and Cross-Validation Evaluation. In this paper we propose a new approach to dealing with financial overfitting, a Covariance-Penalty Correction, in which a risk metric is lowered given the amount of parameters and data used to underpins a trading strategy. We outlined the foundation and main results behind the Covariance-Penalty correction for trading strategies. After that, we pursue an empirical investigation, comparing its performance with some other approaches in the realm of Covariance-Penalties across more than 1300 assets, using Ordinary and Total Least Squares. Our results suggest that Covariance-Penalties are a suitable procedure to avoid Backtesting Overfitting, and Total Least Squares provides superior performance when compared to Ordinary Least Squares. 
	
\end{abstract}



\section{Introduction}

Systematic trading strategies are  rule-based procedures which choose portfolios and  allocate assets, in order to deliver a set of returns which optimize some performance criterion, help to diversify a larger portfolio, or perhaps deliver returns specifically tailored to market conditions, such as  providing insurance during rises of market volatility. In order to attain any of these desired return profiles, quantitative strategists must determine a large array of trading parameters, including determining the universe of assets being held, the appropriate trading horizons, the type of strategy being considered and many other possible parametrizations. Normally the analyst tries to identify the appropriate parameters by using all the historical data available, or by holding-out some of the most recent data in order to use  it as an out-sample validation set. However, both backtesting strategies suffer from drawbacks that can render their findings misleading.

Partly due to the abundance of  misleading results produced from backtests, and partly due to the endemic abuse of backtested results, there is an increasing interest in devising procedures for the assessment and comparison of strategies \cite{Paper:HarveyBacktesting:2015,Paper:RomanoEfficient:2016,Paper:BaileyPBO:2015}. The  goal is to devise schemes for preventing what is known as  {\em backtesting overfitting.} Backtest overfitting can be understood as a collateral effect that occurs during the  identification of strategies,  resembling  other practical modelling problems: overfitting in supervised learning \cite{Book:HastieElements:2016,Proc:DworkHoldout:2015,Proc:KohaviCV:1995}, or data-snooping due to multiple hypothesis testing biases,  (also known as {\em p-hacking}), a  problem  occuring frequently in the applied sciences \cite{Paper:HolmSimple:1979,Paper:BenjaminiFDR:1995}.

So far, many financial researchers have proposed different ways  to tackle this problem that can be broadly categorised in three types: Data Snooping, Overestimated Performance, and Cross-Validation Evaluation. In summary, the Data Snooping and Overestimated Performance techniques aim to identify spurious results by taking into account the number of trials and the performance from all attempted strategies in a multiple hypothesis testing framework. Conversely, Cross-Validation approaches provide more practical and generic mechanisms to individually fine-tune trading strategies.

In this paper we propose a new approach to dealing with financial overfitting, a Covariance-Penalty Correction, in which a risk metric is lowered given the amount of parameters and data used to underpins a trading strategy. We outline the foundation and main results behind the Covariance-Penalty correction for trading strategies. After that, we pursue an empirical investigation, comparing its performance with some other approaches in the realm of Covariance-Penalties across more than 1300 assets, using Ordinary and Total Least Squares -- with the last providing superior performance. Before delving into this novel perspective, we first briefly exhibit a historical context about the available methodologies. 

\section{Literature Review}

We can outline, chronologically, three distinct approaches in the literature to evaluate and deal with backtesting overfitting: Data Snooping, Overestimated Performance, and Cross-Validation Evaluation. The problem of overfitting cannot be understated, and innumerable references highlight the issues with p-hacking which has been an issue for considerable periods but making headlines more recently (see e.g., \cite{Munchhausen}, \cite{Ioannidis}, \cite{OpenScience}, \cite{Freedman}), and \cite{Head})
although it may not always be willful (\cite{Gelman:Forking}), and is pernicious in finance (see for instance, \cite{Suhonen}).

\subsection{Data Snooping}

Data snooping is a problem commonly encountered in the realm of econometrics, specifically when a researcher is exploring asset pricing models. Among several definitions in different contexts, in finance it is determined by the act of finding significant yet spurious risk premia factors, due to the unavoidable reuse of the same time series \cite{Paper:CrackDataSnooping:1999, Paper:WhiteReality:2000}. The first work in this direction can be attributed to Lo and MacKinlay \cite{Paper:LoMacKinlaySnooping:1990} attempted to highlight and quantified the effect of data snooping in risk premia studies. 

More precisely, the data-snooping problems/solutions tend to focus on the issue of Multiple Hypothesis Testing. Standard methods to adjust for snooping from multiple hypothesis testing biases, such as by Bonferroni and Holm to control family wise error rates (FWER\footnote{Probability of making one false discovery, that is, non-rejecting a strategy with Sharpe Ratio $= 0$.}) and by Benjamini-Hochberg-Yekutieli (BHY) for controlling False Discovery Rates (FDR\footnote{Expected frequency of non-rejecting a strategy with Sharpe Ratio $= 0$.}) are in relatively common use in applied sciences \cite{Paper:HolmSimple:1979,Paper:BenjaminiFDR:1995}. These methods adjust for the likelihood of false discoveries due to the correlation between multiple tests and are an effective way to prevent what is otherwise known as p-hacking in the applied sciences. Nonetheless, since they adjust to effectively worst-case scenarios, they are known to result in significant reductions of power.

Due to the low power of the FWER adjustments, Westfall and Young \cite{Book:WestfallResampling:1993} introduced bootstrap methods to estimate the empirical correlation between hypothesis tests as a way of optimising the power of the resulting adjusted test statistics. Following on these, White \cite{Paper:WhiteReality:2000} outlined one of the first attempts to improve test statistics by applying bootstrap methods to identify spurious relationships in the realm of trading strategies, applying them to technical trading rules \cite{Paper:SullivanDangers:2001}. Romano and Wolf \cite{Paper:RomanoExact:2005} advanced in procedures to more adequately correct multiple testing and resourcing in more adequate forms of bootstrapping for standard errors calculations; we point also to more recent works in this direction \cite{Paper:RomanoStepwise:2005, Paper:RomanoFormalized:2008, Paper:RomanoEfficient:2016, Paper:HarveyLucky:2015}. 

Since Romano and Wolf can be seen as an extension of White's work, we briefly outline its algorithm for data snooping diagnosis. Let $r_{it} = R_{it} - R_{bench,t}$ represent the excess return ($R$) of the asset $i$ in relation to a benchmark $bench$ at time $t$. Each asset can represent a trading strategy with different parametrisation, and the benchmark can be the usual 3-month London Interbank Overnight Rate return or simply an offset. Given a differentiable performance function $P$, a significance level $\alpha$, and a number of bootstrap samples $B$ ($b=1,...,B$), do:

\begin{itemize}
	\item \textbf{Compute} each asset performance: $p_i = P(r_{i1}, ..., r_{it}, ..., r_{iT})$. 
	\item \textbf{Set} $\hat{V} = \sqrt{T} * \max(p_1, ..., p_n)$ as the maximum estimated performance across the available assets. 
	\item Set $k=1$, and \textbf{while} all assets have not been rejected:
	\begin{enumerate}
		\item \textbf{For} $b=1,...,B$ bootstrap samples, do:
		\begin{enumerate}
			\item Take a bootstrap sample $r_{.1}^{(b)}, ..., r_{.T}^{(b)}$ from the sequence $r_{.1}, ..., r_{.T}$. The $.$ notation is the abbreviation for "all assets".
			\item \textbf{Compute} each asset performance: $p_i^{(b)} = P(r_{i1}^{(b)}, ..., r_{it}^{(b)}, ..., r_{iT}^{(b)})$
			\item \textbf{Define} the statistic $V_b = \sqrt{T} * \max(p_i^{(b)} - p_1, ..., p_n^{(b)} - p_n)$ 
		\end{enumerate}
		\item \textbf{Find} the critical value $c_k$ by estimating the quantile $(1-\alpha/k)$\% of $V_b$ empirical distribution.
		\item \textbf{Reject} all strategies that $(\sqrt{T} * p_i) > c_k$; remove these and place them in the echelon $k$.
		\item $k+=1$
	\end{enumerate}
	\item \textbf{Report} all echelons to the user.
\end{itemize}

Each echelon represent a subset of trading strategy that tend to perform similarly, but are superior/inferior to the followed/previous iteration. As it can be seen, this method is based on a stepdown approach, since we start with the null hypothesis that correspond to the largest test statistic. A final comment: for the first iteration it is possible to calculate the p-value for the Bootstrap Reality Check \cite{Paper:WhiteReality:2000} by: $p_{BRC} = \frac{\sum_{b=1}^{B} V_b > \hat{V}}{B}$. This statistic addresses the question whether the strategy that appears best in the observed data really beats the benchmark; it does not attempt to identify as many outperforming strategies as possible. We also highlight that this presentation took the non-studentised version of Romano and Wolf procedure; for the studentised version, which provide some improvements to the statistical procedure, refer to \cite{Paper:RomanoExact:2005}.

Finally, another related approached called Model Confidence Set \cite{Paper:HansenMCSVol:2003, Paper:HansenMCS:2011}, acknowledges the limitations of the data, such that uninformative data yield a MCS with many models, whereas informative data yield a MCS with only a few models. By pairwise comparisons, this methodology creates confidence sets (analogous to confidence intervals) that contains the best model or models given a specified significance level.

\subsection{Overestimated Performance}

In Harvey and Liu \cite{Paper:HarveyBacktesting:2015} the authors exhibit a multiple hypothesis testing framework devoted to identify strategies incorrectly classified as profitable. Before delving into the \cite{Paper:HarveyBacktesting:2015} work, it should be mentioned that similar works have also proposed ways to incorporate uncertainty and perform adjustments in the strategies performance (mainly for Sharpe Ratios). A section of these tends to favour hypothesis tests and confidence interval computations, starting from: (i) the classical i.i.d. Gaussian returns \cite{Paper:JobsonPerformance:1981}; (ii) then dropping Gaussian assumption with the support of (independent) bootstrap \cite{Paper:VinodConfidence:1999}; (iii) more ahead being able to pinpoint an asymptotic distribution for the Sharpe Ratio under non-i.i.d and more mild assumptions \cite{Paper:LoSharpeStats:2002, Paper:OpdykeComparing:2007}; and finally (iv) dropping i.i.d. and underlying distribution assumptions by applying improved forms of bootstrapping \cite{Paper:LedoitRobustPerformance:2008}. Another approach is by adjusting the Sharpe Ratio by computing the so-called Probabilistic Sharpe Ratio \cite{Paper:BaileySharpeFrontier:2012}. This measure computes the probability that a strategy performance is greater than a benchmark, taking into account the skewness and kurtosis of the returns.

However, all these works do not take into account the number of trials did to fully develop the strategy, or when a hypothesis test is being made to compare different strategies. This information is relevant, because some improvement is expected whenever the analyst have access to information coming from the backtest: increasing overall performance, removing drawdowns, etc. In this way, he can design stop/loss rules, change leveraging rules, with undoubtly impact in the strategy performance when the backtest is remade. The authors \cite{Paper:HarveyBacktesting:2015} treat this as a Multiple Hypothesis Testing situation \cite{Paper:HolmSimple:1979, Paper:BenjaminiFDR:1995}. To illustrate this concept, if we suppose that the (excess) returns are i.i.d. and follows a Normal distribution, then the Sharpe Ratio can be reinterpreted as t-scores:
\begin{equation}
t_{score} = \frac{\bar{R}}{\sigma_{R}} \times \sqrt{T} = \hat{SR} \times \sqrt{T}
\end{equation}

where $\hat{SR}$ is the estimated Sharpe Ratio and $T$ is the backtest length. We can compute a t-test using this $t_{score}$, aiming to check the probability for observing a departure from the null hypothesis ($H_0: SR=0$) as extreme as the difference present in our sample data. This is define as:
\begin{equation}
p^{S} = Pr(|t| > t_{score}) \label{singletest}
\end{equation}

in which $t$ denotes a random variable that follows a t-distribution with $T-1$ degrees of freedom, and $p^{S}$ is commonly dubbed as p-value, which in this case is computed for a single hypothesis ($S$). If a research have explored several strategies, then some overstatement will appear in the $\hat{SR}$, and we could interpret this as the researcher did tacitly multiple attempts to obtain such result. As a initial estimation of this overestatement, suppose that all attempts are independent (unreal, but we drop this more ahead), then we can rewrite \ref{singletest} as:
\begin{eqnarray}
p^{M} = Pr(max(|t_i|, i=1, ..., n) > t_{score}) = \nonumber \\ = 1 - \prod_{i=1}^n Pr(|t_i| \leq t_{score}) = 1 - (1-p^S)^n \label{multitest}
\end{eqnarray}

by following \cite{Paper:HarveyBacktesting:2015}: when $n = 1$ (single test) and $p^S = 0.05$, $p^M = 0.05$, so there is no multiple testing adjustment. If $n = 10$ and we observe a strategy with $p^S = 0.05$, $p^M = 0.401$, implying a probability of about 40\% in finding an investment strategy that generates a $SR$ that is at least as large as the observed $\hat{SR}$, but much larger than the 5\% probability for a single test. It is clear that multiple testing greatly reduces the statistical significance of single test. Hence, $p^M$ is the adjusted p-value, that reflects the likelihood of finding a strategy that is at least as profitable as the observed strategy after $n$ attempts. By equating the p-value of a single test to $p^M$, we obtain the defining equation for the multiple testing adjusted (haircut) Sharpe ratio $\hat{HSR}$:
\begin{equation}
p^M = Pr(|t| > \hat{HSR} \times \sqrt{T})
\end{equation}

since $p^M$ is larger than $p^S$, $\hat{HSR}$ will be smaller than $\hat{SR}$. However, when the test statistics are dependent, however, the approach in the example is no longer applicable as $p^M$ generally depends on the joint distribution of the $n$ test statistics. For this more realistic case, the authors apply techniques that controls the FWER and FDR. These methods are Bonferroni and Holm p-value corrections to control FWER \cite{Paper:DemvsarStatsComp:2006} and BHY procedure for FDR \cite{Paper:BenjaminiFDR:1995}. Below, in Figure \ref{Haircut} we display some charts with pre and post-adjustment of Sharpe Ratios, as well as the Haircuts\footnote{In order to facilitate researches interested on using this approach, they provide a link for the publicly available Matlab code (\url{http://faculty/fuqua.duke.edu/~charvey/backtesting}). It is also possible to apply the same procedure for the Value-at-Risk metric (and therefore, making the same assumptions -- check endnote 23 from \cite{Paper:HarveyBacktesting:2015}).}:

\begin{figure}
	\centering
	\includegraphics[scale=.75]{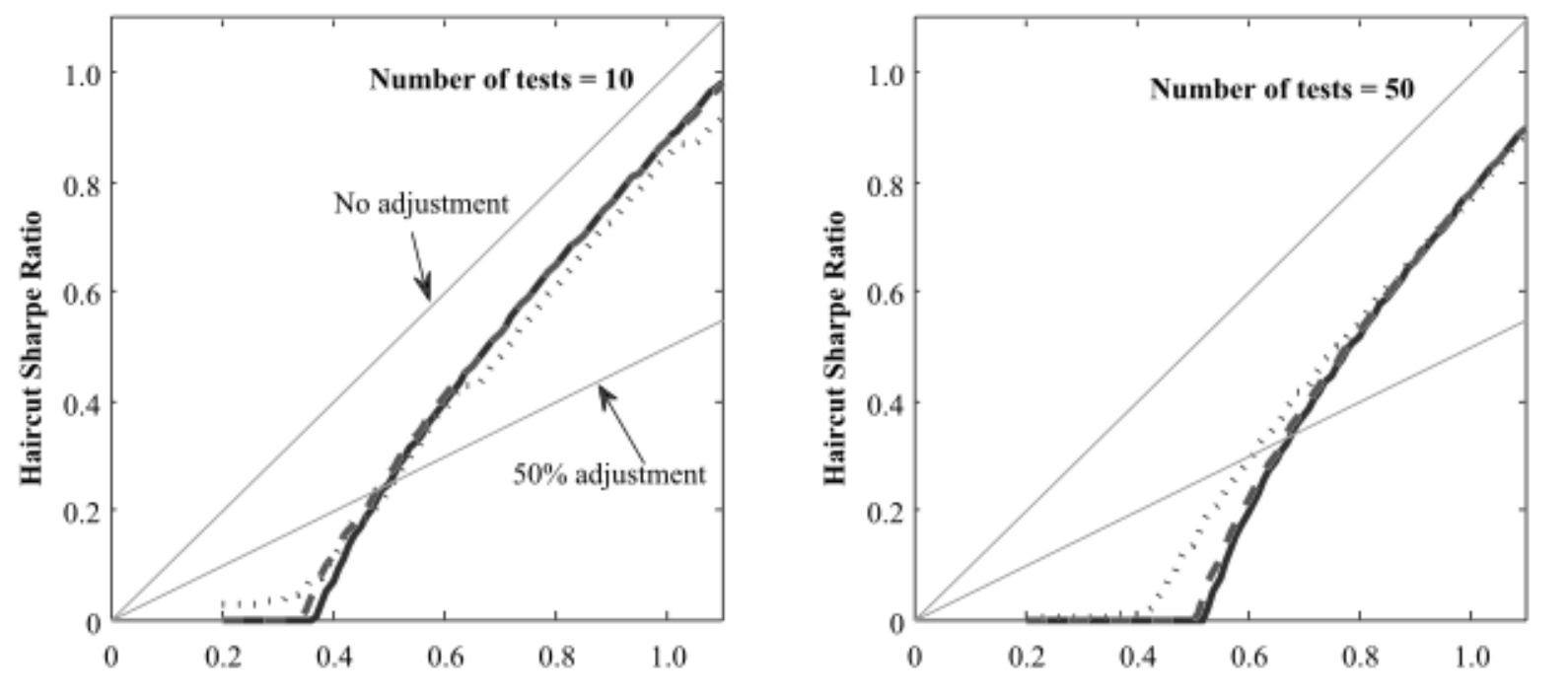}
	\includegraphics[scale=.7]{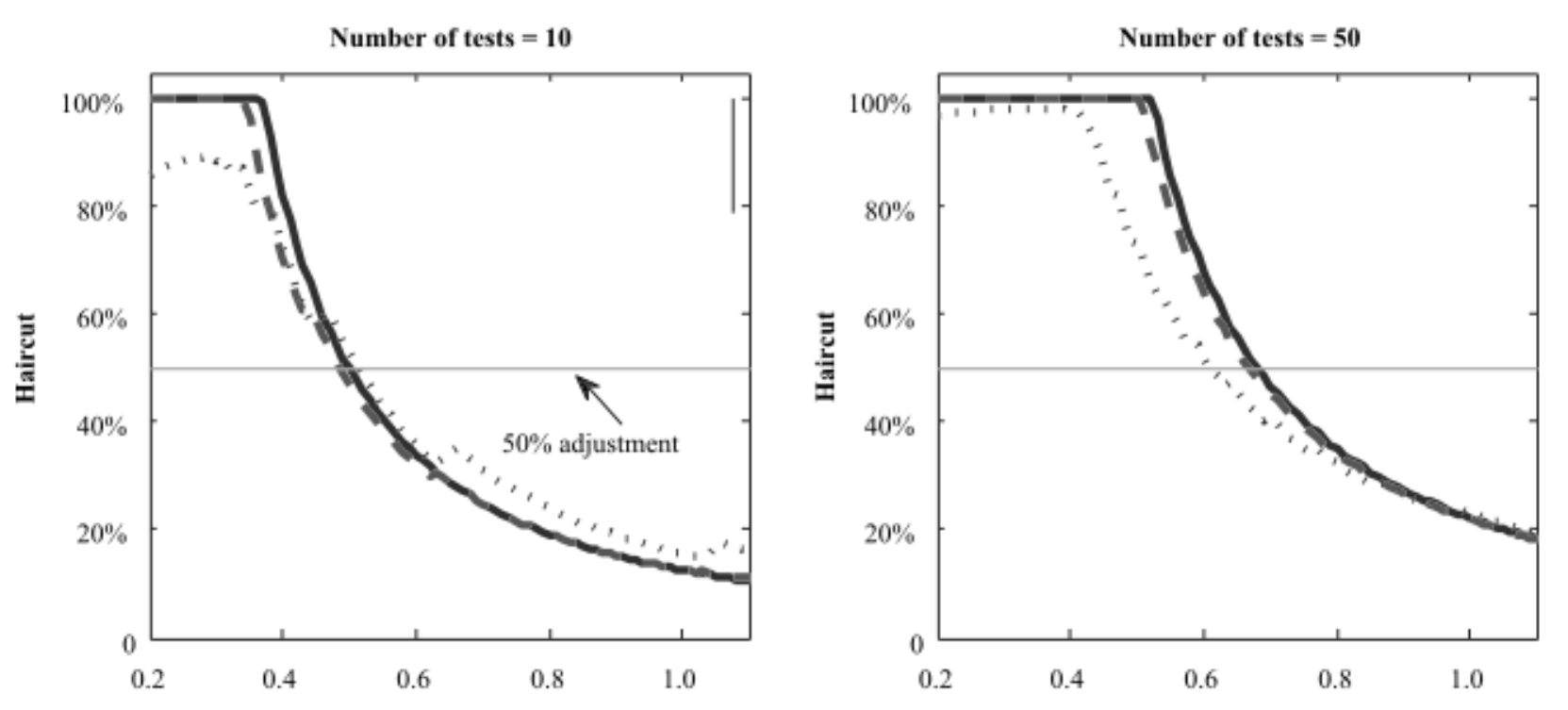}
	\caption{Original and Haircut Sharpe Ratios, as well as the Haircut levels for different methodologies and number of trials. Source: \cite{Paper:HarveyBacktesting:2015}} \label{Haircut}
\end{figure}

As it can be seen, the Haircut is a non-linear function and is highly affected by the number of trials, and the method of adjustment. The authors advocate by the BHY method in financial applications, as they argue that it is more reasonable to control for the rate of false discoveries, rather than controlling for not making any false discovery. Finally, as the authors points out, the main caveat of this method is that all the analysis is based with in-sample data, and as they observe a hybrid strategy might be useful as a double validation of the selected trading strategy.

In line with multiple testing literature, there remain challenges associated with knowing which analyst observations of data are, in fact, uses of the data and thus, tests. Other challenges of multiple testing (in particular, when using brute-force penalisations such as BHY, Holm and Bonferroni) include:
\begin{itemize}
	\item What is the impact of common knowledge? 
	\item How does prior information impact hypotheses being tested? (and strategies being designed)?
	\item If multiple analysts look at the same data, should they be penalized based on each other observations?
	\item If the tests were conducted in a different order, the results would be different, so how does this impact the philosophy of the approach?
	\item How do we know if we have been influenced by others who have seen the same data?
	\item Is it {\em even} possible to have no knowledge of relevant financial history? 
\end{itemize}
While care and concern over the overuse of history is imperative, it is not altogether clear whether the philosophy of multiple testing, especially as done via BHY, Holm and Bonferroni, is reasonable for financial applications. Nonetheless, we believe the methodology is absolutely informative and can be used in conjunction with our own. As yet an unresolved area, preventing overfitting requires a multiplicity of approaches.

\subsection{Cross-Validation Evaluation}

In Bailey et al. \cite{Paper:BaileyPBO:2015} the authors provide a set of definitions, tools and benchmarks to characterize the so-called backtest overfitting. In certain sense this work extend previous research and findings from the same group of authors, such as the probability that a Sharpe Ratio is inflated \cite{Paper:BaileySharpeFrontier:2012} and to determine the minimum track record for a Sharpe Ratio to be statistically different from a given threshold \cite{Paper:BaileyPseudo:2014}.

In terms of definitions, the authors establish that a strategy selection overfits if the expected performance in-sample is less than the median performance rank out-of-sample of all strategies. By strategy selection, it means a way of ranking the available strategies using a set of performance metrics (e.g., Sharpe Ratio). 

\begin{figure}
	\centering
	\includegraphics[scale=0.6]{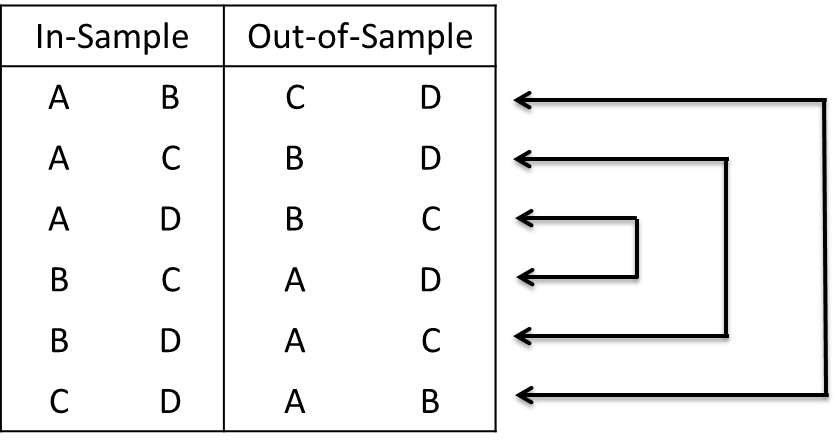}
	\caption{Generating the CSCV folders, when $S=4$. Each letter represents a block, and note that whenever the shuffles happen, it is necessary to reorder the blocks in each in-sample and out-of-sample sets.} \label{CSCV}
\end{figure}

In relation to tools they propose the procedure called Combinatorially-Symmetric Cross-Validation (CSCV), which is briefly outlined below and represented in Figure \ref{CSCV}.
\begin{enumerate}
	\item Split the profit and loss of several strategies in $S$ even size blocks over the backtested horizon (where $S$ is a even natural number);
	\item Reshuffle these blocks in all feasible arrangements such that the natural ordering is as most preserved as possible; and
	\item For each first half (in-sample) of the arrangement, find the optimal strategy for a given performance measure and compare it to its own ranking in the out-of-sample set (second half).
\end{enumerate}

With these rankings and comparisons, the authors compute four complementary metrics:

\begin{enumerate}
	\item Probability of Backtest Overfitting: the probability that the model configuration selected as optimal in-sample will underperform the median of the $N$ models configurations out-of-sample.
	
	\item Performance degradation: this determines to what extent greater performance in-sample leads to lower performance out-of-sample (discussed in \cite{Paper:BaileyPseudo:2014}).
	
	\item Probability of loss: the probability that the model selected as optimal IS will deliver a loss OOS.
	
	\item Stochastic dominance: this analysis determines whether the procedure used to select a strategy IS is preferable to randomly choosing one model configuration among the $N$ alternatives.
\end{enumerate}

Also, in terms of benchmarks the authors propose some synthetic and practical test cases. In this sense, the objective of these studies is to assess if the proposed framework is able to dodge from overfitted strategies with different set of hyperparameters (number of strategies, backtest length and overfit degree). 

We can highlight two main critics assessment towards this work: 

\begin{enumerate}
	\item Preservation of temporal dependence: the authors claim that their method preserves the temporal dependence. In their support are the fact that their method reorders the blocks in-sample and out-of-sample and someone might argue that daily returns tends to present weak autocorrelation. However, this claim about preservation of time dependence is clearly not absolutely true: just check the third to fifth row at Figure \ref{CSCV}.
	
	\item High overlap degree in the in-sample and out-of-sample metrics: it can be easily noticed that there is a high level of juxtaposition of blocks appearing several times in the out-of-sample part (and similarly in the in-sample). The authors in proposing the CSCV, using as inspiration the k-fold cross-validation, forgot to analyse deeper the impact of the highly correlated out-of-sample folders (something that do not occur in the case for the k-fold cross-validation). This correlation exerts an impact in the computation of standard errors and hypothesis tests \cite{Proc:KohaviCV:1995, Paper:BengioCV:2004}, and therefore decrease the out-of-sample assessment quality.
\end{enumerate}

\subsection{Summary}

Table \ref{summary:backrev} present a categorisation of all the previous lines of investigation -- with the addition of Covariance-Penalty --, in terms of their breadth/generality of use in practice and the type of solution used to tackle backtesting overfitting. 

\begin{table}[h!]
	\centering
	\caption{Categorisation of all the four lines of investigations in backtesting overfitting.} \label{summary:backrev}
	\begin{tabular}{l|c|c}
		\hline
		\hline
		Breadth / Solution & Multiple Hypothesis & In-Out Sample Assessment \\
		\hline
		Generic & Data Snooping & Cross-Validation \\
		\hline
		Assumptions & Overestimated Performance & Covariance-Penalty \\
		\hline
		\hline
	\end{tabular}
\end{table}

The Data Snooping and Overestimated Performance authors identify that the root cause of any overfitting or spurious results is due to the multiple trials (hypothesis) employed by the analyst during the strategies fine-tuning step. However, the Data Snooping solutions tend to be more generic -- it is based on bootstrapping returns --, with the expense of a increased computational burden. The Overestimated Performance solutions are based on a set of assumptions, allowing the authors to find closed-form solutions or simpler algorithms that make the performance correction aspect easier to compute. 

In contrast, the Cross-Validation Evaluation and Covariance-Penalty Correction are focused on avoiding the apparent in-sample performance, by uncovering the expected generalisation (out-sample) potential of a strategy. Similarly, they divert in terms of their amplitude: the Covariance-Penalty Correction methods require some assumptions on the joint behaviour of the asset and trading strategy, which is not the case for the majority of the Cross-Validation procedures. However, this lack of assumptions demand an extra computational processing in respect to the Covariance-Penalty approach.

%\subsection{Other Alternative Strategies}
%Other strategies for dealing with overfitting have been proposed including the use of Hierarchical Bayesian models (\cite{Gelman2012}) using combined priors on the various models (imposing a population effect rather than individual effects). This method effectively imposes a bias towards more uniform estimates on the various individual effects.
%
%Other methods include the use of {\em Differential Privacy} from the information security literature as a way of doing out-of-sample or holdout testing. Statisticians are allowed to see test and train sets, but should only visit the holdout set once, ever in order to ensure that the resulting statistics are untainted by multiple-testing biases. Differential privacy based methods do not allow researchers to use the holdout dataset but they instead can see a sample statistic computed from it. Were they able to find the statistic enough times, they would be able to reveal the data from the holdout. Differential privacy-based methods prevent this holdout snooping by effectively noising the goodness-of-fit statistics. These methods or their underlying statistical basis may prove a promising approach to overfitting problems.
%
%- a carefully constructed randomized sample to ensure non-invertible tests. This is closely related to work by Blum-Hardt on Kaggle-competition hacking.
%* New paper by Google, MS, IBM, Samsung, et al authors \cite{Dwork1} \cite{Dwork2} \cite{Dwork3}

\section{Covariance-Penalty for Trading Strategies}

This section outline the theoretical results and notations used along this paper. We start by enunciating single period linear strategies, and some of its properties. Then we move to optimisation of trading strategies by maximising correlation, and its links to Sharpe ratio utility function. This result/link allows us to conclude this section with the main result of this paper: Covariance-Penalty correction of Sharpe ratio.

\subsection{Single period linear strategies}

We consider the (log) returns of a single asset, ${R_{t}\sim\mathscr{N}(0,\sigma_R^{2})}$ returns with auto-covariance function at lag $k$, $\gamma(k)=E[R_t R_{t-k}]$, together with corresponding auto-correlation function (ACF), $c(k)=\gamma(k)/\gamma(0)$ at lag $k$. 

Our main aim is to work with strategies based on linear portfolio weights (or {\sl signals}) $X_{t}=\Sigma_{1}^{\infty}a_{k}R_{t-k}$ for coefficients $a_{k}$ generating the corresponding dynamic strategy returns $S_t=X_t\cdot R_t$ (here, and always, the signal, $X_t$ is assumed to only have appropriately lagged information).
Example strategy weights include exponentially weighted moving averages $a_{k}\propto\lambda^{k}$, simple moving averages $a_k = \frac{1}{T} \mathbbm{1}_{[1,\ldots,T]}$,  forecasts from ARMA models, etc. Most importantly, the  portfolio weights $X$ are normal and {\em jointly} normal with returns $R$.

We restrict our attention to return distributions over a single period. In the case of many momentum strategies, this period can be one day, if not longer. For higher-frequency intra-day strategies, this period can be much shorter. The pertinent concern is that the horizon (i.e., one period) is the same horizon over which the rebalancing of strategy weights is done. If weights are rebalanced every five minutes, then the single period should be five minutes. This is a necessary assumption in order to ensure the joint normality of (as yet indeterminate) signals and future returns. Moreover, this assumption will give some context to our results, which imply a maximal Sharpe ratio, maximal skewness and maximal kurtosis for dynamic linear strategies. 

We are interested in characterizing the moments of the strategy's unconditional returns, the corresponding standard errors on estimated quantities, and means of optimising various non-dimensional measures of returns such as the Sharpe ratio via the use of non-linear transformations of signals. Our goal is to look at unconditional properties of the strategy. It is important to avoid foresight in strategy design and this directly impacts the  conditional properties of strategies (e.g., conditional densities involve conditioning on the currently observed signal to determine properties of the returns, which are just Gaussian). In the context of our study, we are concerned with  one-period ahead returns of the   unconditional returns distribution of our strategy, where both the signals and the returns are unobserved, and the resulting distributions (in our case, the product of two normals) are much richer and more realistic.

\subsection{Properties of linear strategies}

Given the joint normality of the signal and the returns, we can explicitly
characterise the one-period strategy returns (see \cite{Exact}). To allow for greater extendibility, we prefer to only consider the moments of the resulting distributions. These can be characterized easily using  Isserlis' theorem \cite{Isserlis}, which gives all moments for any multivariate normal random variable in terms of the mean and variance.  We also refer to  \cite{Haldane} who meticulously produces both non-central and central moments for powers and products of Gaussians. While this is a routine application of Isserlis' theorem, the algebra can be tedious, so we quote the results. 

%\greybox{
%\vskip\baselineskip
\begin{theorem}[Isserlis (1918) \cite{Isserlis}]
	If $X \sim \mathscr{N}(0,\Sigma)$,then 
	$$E[X_1 X_2\cdots X_{2n}] = \sum_{i=1}^{2n} \prod_{i \neq j} E[X_i X_j]$$

	and 
	$$E[X_1 X_2\cdots X_{2n-1}] = 0$$
	where the $\sum\prod $ is over all the $(2n)!/(2^n n!)$ unique partitions of $X_1,X_2,\ldots X_{2n}$ into pairs $X_i X_j$.
\end{theorem}

Haldane's paper quotes a large number of moment-based results for various powers of each normal. We quote the relevant results.

\begin{theorem}[Haldane (1942) \cite{Haldane}]
	\label{thm:Haldane}
	If $x,y\sim \mathscr{N}(0,1)$ with correlation $\rho$ then 
	\begin{eqnarray}
		E[xy] =& \rho \nonumber \\ 
		E[x^2y^2]=&1+2\rho^2 \nonumber \\
		E[x^3y^3]=&3\rho(3+2\rho^2) \nonumber \\
		E[x^4y^4] =&3(3+24\rho^2+8\rho^4) \nonumber
	\end{eqnarray}
	and thus the central moments of $xy$ are
	\begin{eqnarray}
	\label{eqn:centered_moments1}
	\mu_1 =& \rho \nonumber \\
	\label{eqn:centered_moments2}
	\mu_2=&1+\rho^2 \nonumber \\ 
	\label{eqn:centered_moments3}
	\mu_3=&2\rho(3+\rho^2) \nonumber \\
	\label{eqn:centered_moments4}
	\mu_4= &3(3+14\rho^2+3\rho^4) \nonumber
	\end{eqnarray}
\end{theorem}
%\vskip\baselineskip
%}  End of greybox

From these one period moments, (and a simple scaling argument giving the dependence on $\sigma(x)$ and $\sigma(y)$) we can characterise Sharpe ratio, skewness, etc., and can also define objective functions in order to determine some sense of optimality for a given strategy.

\begin{theorem}[Linear Gaussian]
	\label{thm:Linear}
	For single asset returns and a one period strategy, $\ensuremath{R_{t}\sim\mathscr{N}(0,\sigma_R^{2})}$  and $X_t\sim\mathscr{N}(0,\sigma_X^2)$ jointly normal with correlation $\rho$, the Sharpe ratio is given by
	\begin{equation}
		SR = {\rho \over \sqrt{1+\rho^2}} \label{Eqn:Sharpe}
	\end{equation}
	the skewness is given as
	\begin{equation}
	\gamma_3=\frac{2\rho(3+\rho^{2})}{(1+\rho^{2})^{\frac{3}{2}}}, \label{Eqn:Skew}
	\end{equation}
	and the kurtosis is given by
	\begin{equation}
	\gamma_4 = \frac{3(3+14\rho^2+3\rho^4)}{(1+\rho^2)^2}
	\label{Eqn:Kurt}
	\end{equation}
\end{theorem}
In the appendix, we extend  equations (\ref{Eqn:Sharpe}) and (\ref{Eqn:Skew})   to the case of non-zero means.

% {\bf Proof:  } 
\begin{proof}
	A simple application of  Theorem \ref{thm:Haldane} give us the following first two moments for our strategy $S_t = X_t \cdot R_t$:
	$\mu_1=E[S_t]=E[X\cdot R] = \sigma_X\sigma_R \rho $.
	and $\mu_2=Var[S_t]= \sigma_X^2\sigma_R^2(\rho^2+1)$
	.
	Thus we can derive the following results for the Sharpe ratio,
	\begin{eqnarray}
		SR = {\mu_1\over \mu_2^{1/2}} = {\sigma_X\sigma_R\rho\over \sigma_X\sigma_R \sqrt{\rho^2+1}} = {\rho\over  \sqrt{\rho^2+1}}
	\end{eqnarray} 
	Moreover, we can see that the skewness,
	\begin{eqnarray}
		\gamma_3 =  {\mu_3\over \mu_2^{3/2}} =  {2\rho (3+\rho^2)\over (1+\rho^2)^{3/2}}
	\end{eqnarray}
	Finally, the kurtosis is given by
	\begin{eqnarray}
		\gamma_4 = {\mu_4\over \mu_2^{2}} = {3(3+14\rho^2+3\rho^4)\over (1+\rho^2)^2}
	\end{eqnarray}
	% $\Box$
\end{proof}

\begin{figure}[h]
	\includegraphics[width=\linewidth]{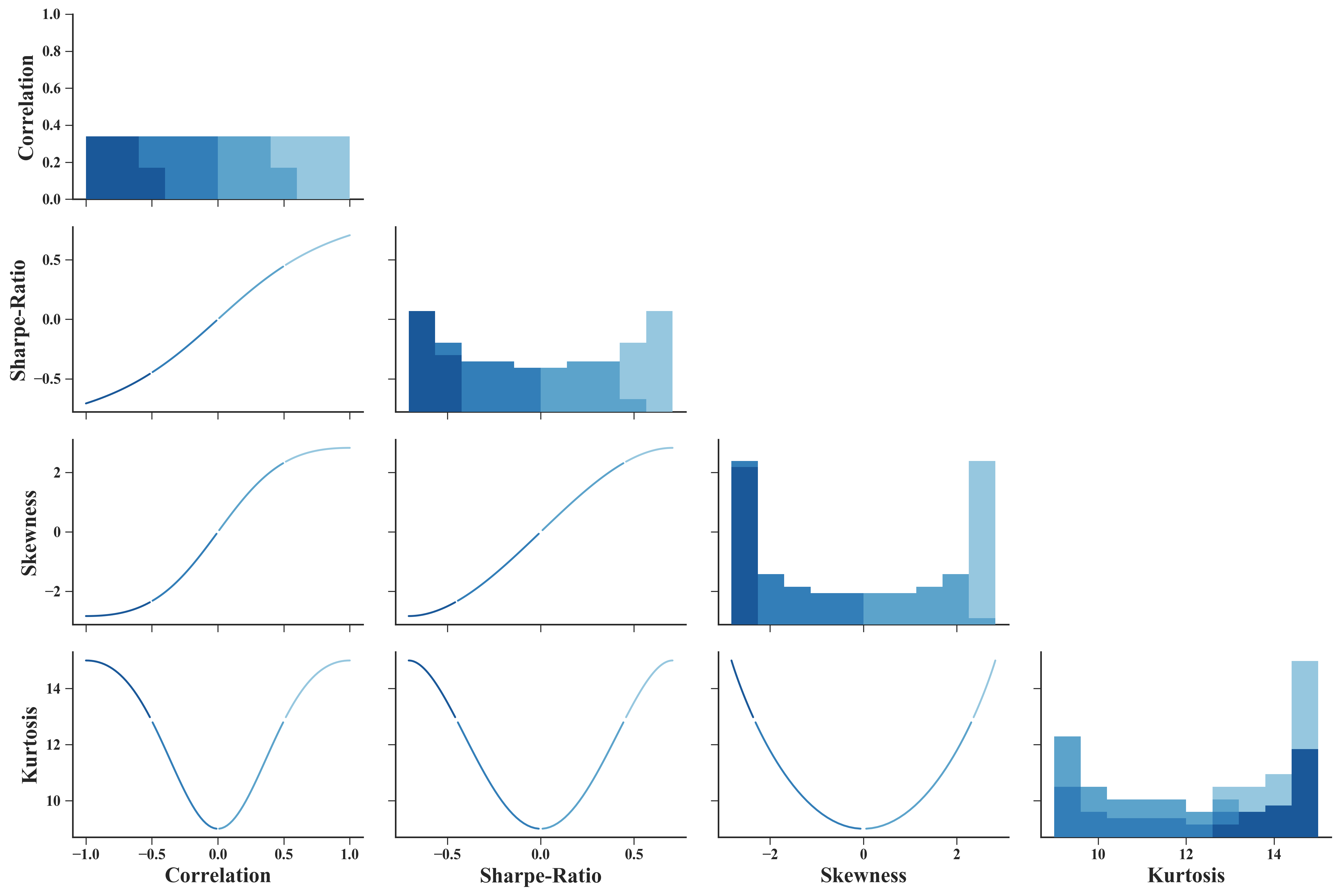}
	\caption{{\bf Correlation, Sharpe ratio, Skewness, and Kurtosis pairwise relationship.} A uniform distribution in correlation is bucketed into four ranges $\{[-1,-0.5],[-0.5,0],[0,0.5],\ [0.5,1]\}$ as depicted in the bar charts in shades of blue. After transforming the correlation into SR, $\gamma_3$ and $\gamma_4$ the frequencies are no longer uniform.} \label{fig:SharpeSkew}
\end{figure}

If we restrict our attention to positive correlations, all three dimensionless statistics are monotonically increasing in $\rho$. Consequently, strategies that maximize one of these statistics will maximize the others, although the impact of correlation upon Sharpe ratio, skewness and kurtosis is different. We illustrate the cross-dependencies in the following charts, depicting the relationships between the variables. In figure \ref{fig:SharpeSkew}, the shaded blue histograms correspond to correlation ranges ($\{[-1,-0.5],[-0.5,0],[0,0.5],\ [0.5,1]\}$). We note that a uniform distribution in correlations maps into a higher likelihood of extreme Sharpe ratios and an even higher likelihood of extreme skewness and kurtosis. 

Skewness ranges in $[-2^{3/2},2^{3/2}]\approx[-2.8,2.8]$. Unlike the Sharpe ratio,
Skewness' dependence on correlation tends to flatten, so to achieve
90\% peak skewness, one needs only achieve a 0.60 correlation, while
for a 90\% peak Sharpe, one needs a correlation of 0.85. Kurtosis is an even function and varies from a minimal value of 9 to a maximum of 15. In practice, correlations will largely be close to zero and the resulting skewness and kurtosis significantly smaller than the maximal values.

Although we analyse  the moments of the strategy $S_t=X_t R_t$, the full product density is actually known in closed form (see appendix \ref{Section:FullDist},  \cite{Exact} and \cite{DistCorrel}). It is clear that the distribution of the strategy is {\sl leptokurtic} even when it is not predictive (when  the correlation is exactly zero, the strategy has a kurtosis of $9$). In the limit as $\rho\rightarrow 1$, the strategy's density approaches that of a non-central $\chi^2$, an effective {\sl best-case} density when considering the design of optimal linear dynamic strategies.

An optimised strategy with sufficient lags (and a means of ensuring parsimony) may be able to capture both mean-reversion and trend and result in yet higher correlations. Annualised Sharpe ratios of between 0.5-1.5 are most common (i.e., correlations of between 3\% to 9\%) for single asset strategies in this relatively low-frequency regime.

% As shown in figure (\ref{fig:BivDensityChartModels}), the strategies depicted  all have correlations of between 3\%-6\%, or Sharpe ratios via our plug-in formula of between 0.48-0.95 when annualised (ignoring transaction costs which are generally quite considerable for reversal strategies which can have  average daily turnover of about 30\% net change in position!). These effectively range from passable  to modestly better strategies, although there is obviously further room to improve them.

\subsection{Optimisation: Maximal Correlation, Total least squares}

% MOVE THIS TO THE INTRO. RESTRUCTURE
% #TODO

Many algorithmic traders will explain how problematic  strategy optimisation is, given the  endless concerns of over-fitting, etc. Although these are a concern, the na\"{\i}ve use of strategies which are merely {\sl pulled out of thin air} is equally problematic, where there is no explicit use of optimisation (and, in its place more  {\em eye-balling} strategies or targeting Sharpe ratios rather loosely, effectively a somewhat loose mental optimisation exercise).  Practical considerations abound and real-world returns are neither Gaussian nor stationary. We argue irrespectively that using optimisation and a well-specified utility function as a starting point is a means of preventing strategies from being  just untested heuristics.  Unlike most discretionary traders' heuristics  (or {\em rules of thumb}) which have their 
place as a means of dealing with uncertainty (see for example \cite{Gigerenzer}), 
heuristic quantitative trading strategies run the risk of being entirely arbitrary, or are subject to a large number of human biases, in marked contrast to the monniker {\em quantitative} investment strategies.

\begin{figure}[h!]
	\includegraphics[width=\linewidth]{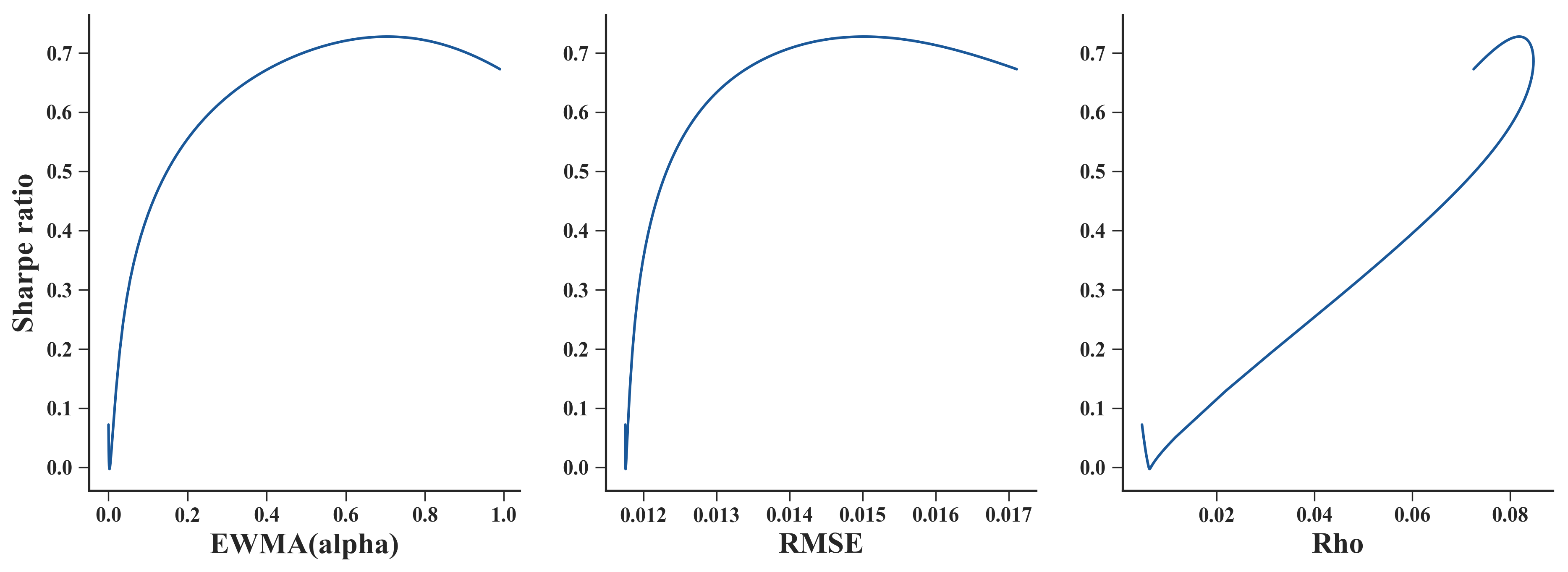}
	\caption{{\bf EWMA Strategy Sharpe Ratio vs $\alpha$, MSE and correlation} for S\&P 500 reversal strategies} \label{fig:EWMA_Sharpe_Corr}
\end{figure}

\begin{figure}[h!]
	\includegraphics[width=\linewidth]{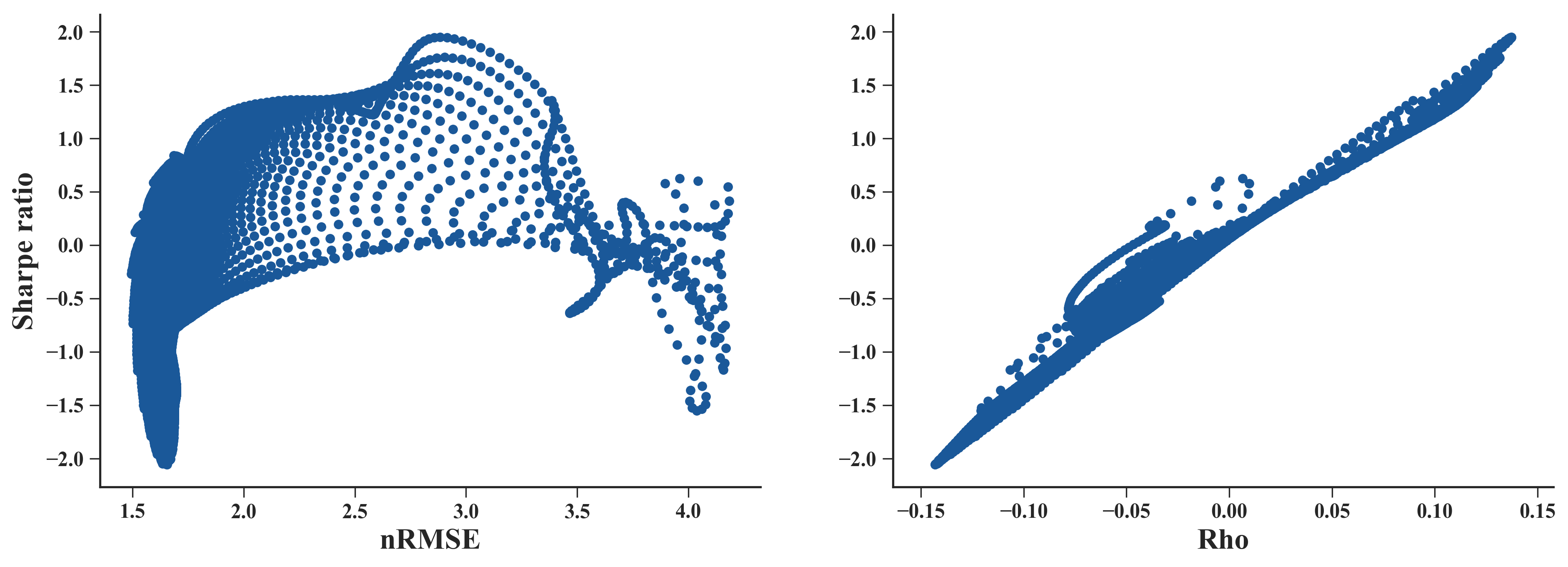}
	\caption{{\bf Holt-Winters Strategy Sharpe Ratio vs MSE and correlation} for S\&P 500 Reversal Strategies} \label{fig:HW_Sharpe_Corr}
\end{figure}

Where optimisation is used, the most common optimisation method is to minimize the mean-squared error (MSE) of the forecast. Our results show that  rather than to minimize the $\mathcal{L}^2$ norm between our signal and the forecast returns (or to maximize the likelihood), if the objective is to  maximize the Sharpe ratio, we must maximize the correlation. 

We can see in figures \ref{fig:EWMA_Sharpe_Corr} and \ref{fig:HW_Sharpe_Corr}, a depiction of fits of strategies applied to S\&P 500 using EWMA and HW filters for a variety of parameters. The relationship between MSE and Sharpe ratio is not monotone in MSE for the EWMA filter as we see in figure \ref{fig:EWMA_Sharpe_Corr}, while it is much closer to being linear in the case of the relationship between correlation and Sharpe. For the case of HW (with two parameters), in figure  \ref{fig:HW_Sharpe_Corr} any given MSE can lead to a non-unique Sharpe ratio, sometimes with a very broad range, leading us to conclude that the optimization is poorly posed. The relationship of correlation to Sharpe is obviously closer to being linear, with higher correlations almost always leading to higher Sharpe ratios.

In the case of a one-dimensional forecasting problem with (unconstrained) linear signals, optimizing the correlation amounts to using  what is known as {\sl total least squares regression (TLS)} or {\sl orthogonal distance regression}, a form of principal components regression (see, e.g., \cite{Golub} and \cite{Markovsky}).  In the multivariate case, it would be more closely related to {\sl canonical correlation analysis} (CCA). 

Unlike OLS, where the dependent variable is assumed to be measured with error and the independent variables are assumed to be measured without error, in total least squares regression, both dependent and independent variables are assumed to be measured with error, and the objective function compensates for this by minimizing the sum squared of orthogonal distances to the fitted hyperplane. This is a simple form of errors-in-variables (EIV) regression and has been studied since the late 1870s, and is most closely related to principal components analysis. For $k$ regressors, the TLS fit will produce weights which are orthogonal to the first $k-1$ principal components. 

So, if we consider the signal $X=Z\beta$ to be a linear combination of features, with $Z\in \mathbf{R}^k$ a $k$-dimensional feature space, then we note that
$$\hat{\beta}^{OLS} = (Z'Z)^{-1}Z'R$$ but
$$\hat{\beta}^{TLS} = (Z'Z-\sigma_{k+1}^2 I)^{-1}Z'R$$ where $\sigma_{k+1}$ is the smallest singular value for the $T\times (k+1)$ dimensional matrix $\tilde{X}=[R,Z]$ (i.e., the concatenation of the features and the returns, see, e.g., \cite{Rahman-Yu}\footnote[4]{A more common method for extracting TLS estimates is via a PCA of the concatenation matrix $\tilde{X}$, where $\hat{\beta}^{TLS}$ is chosen to cancel the least significant principal component}).It is well known that, for the case of OLS, the smooth or hat matrix $\hat{R} = M R$ is given by
$$M^{OLS} = Z(Z'Z)^{-1} Z'$$
with $Tr(M^{OLS})=k$, the number of features. 
In contrast, 
$$M^{TLS}=Z(Z'Z-\sigma_{k+1}^2 I)^{-1}Z'$$
and  effectively has a greater number of degrees of freedom than that of OLS, i.e., $$Tr(M^{TLS})\geq Tr(M^{OLS})$$ with equality only when there is complete collinearity\footnote[5]{In this case, it is also known that $Tr(M)=Tr(L)$ where $L= (Z'Z-\sigma_{k+1}^2 I)^{-1}Z'Z$ and we know that the singular values of $\sigma(L)=\{{\lambda_i^2}/{(\lambda_i^2-\sigma_{k+1}^2)}\}$ where $\lambda_i$ are the singular values of $Z$ (or correspondingly, $\lambda_i^2$ are the singular values of $Z'Z$), and $\lambda_1\geq\cdots\geq\lambda_k>0$ (\cite{Leyang}).  By the Wilkinson interlacing theorem, $\lambda_k\geq\sigma_{k+1}\geq0$ (see \cite{Rahman-Yu}). Consequently,
	$$Tr(M^{TLS})=\sum_i\frac{\lambda_i^2}{(\lambda_i^2-\sigma_{k+1}^2)}\geq k=Tr(M^{OLS})$$ 
	with equality iff $\sigma_{k+1}^2=0$ (i.e., when there the $R^2=100\%$ and consequently, OLS and TLS coincide). In other words,  $Tr(M^{TLS}) \geq Tr(M^{OLS})$.  
}
For this reason, many people see TLS as  an {\em anti-regularisation} method and may result in less-stable response to outliers (see for example, \cite{Zhang}).  Consequently, there is extensive study of {\em regularised} TLS, typically using a weighted ridge-regression (or Tikhonov) penalty (see discussion in \cite{Zhang} for more detail on this large body of research).

%% Yes Cassella-Berger does do TLS. I think some of the best papers on it are by numerical analysts
%% http://people.duke.edu/~hpgavin/SystemID/CourseNotes/TotalLeastSquares.pdf
%% ftp://ftp.esat.kuleuven.ac.be/pub/stadius/ida/reports/91-35.pdf

While maximizing correlation rather than minimizing the MSE seems a very minor change in objective function, the formulas differ from those of standard OLS. The end result is a linear fit which takes into account the errors in the underlying conditioning information. We believe that it should be of relatively little  consequence when the features are appropriately normalized, as is the case for univariate time-series estimation, although some authors have suggested that optimising TLS is not appropriate for prediction (see, e.g., \cite{Fuller} section 1.6.3).  When we seek to maximize the Sharpe ratio of a strategy, the objective should {\em not be} prediction, but rather optimal weight choice.

% * <as.koshiyama@gmail.com> 2017-08-21T19:55:32.617Z:
% 

\subsection{Covariance-Penalty Correction}

\label{Overfitting}

In order to partly address the issue of overfitting of dynamic models, we look to estimated objective functions, taking degrees of freedom into account. Most commonly used estimates include Akaike and Bayes' information criteria, but these require the use of a likelihood. In our case, we do not have an exact likelihood, but instead choose to optimise the Sharpe ratio, and noting the monotonicity in $\rho$, this comes from maximizing $\hat{\rho}$. Consequently, we wish to consider the means of optimising correlation taking into account penalisations for degrees of freedom. This type of penalised optimum can be found via Mallows' $C_p$ or via Cross validation.

We mirror the discussion in \cite{CASI} on Covariance penalties, Section 12.3, which unifies the discussion of Mallow's $C_p$, Stein's Unbiased Risk Estimate (SURE), Akaike (AIC), and Bayesian Information Criteria (BIC). Each of these are methods for adjusting estimated in-sample goodness of fits measures, to produce better estimates of the out-of-sample performance. Effectively, they are a  means of adjusting for the degrees of freedom of various models, akin to adjusted $R^2$ measures.

We are particularly interested in the biases which occur due to in-sample estimates of measures which we optimize. Standard errors will seem lower in-sample, due to the fact that we have simultaneously optimized correlation, and measured this optimum to produce an estimated Sharpe ratio . We would like to have a more sober estimate of the potential performance of our optimised strategies out of sample.

In our study, we do not fit probabilistic models or data-generating processes, but rather  optimize the strategy. As a consequence, we do not have explicit MLE estimators  (or Bayesian posterior), and no likelihood is explicitly used, so we cannot reproduce results for the AIC and the BIC. We are, however, able to find estimates for optimal $\hat{\rho}_{in}$ and the resulting out-of-sample adjustments for this $\hat{\rho}_{in}$ which give us a more reasonable expectation of the real $E[\rho_{out}]$ out of sample\footnote[5]{To avoid confusion, we distinguish between our in-sample estimator $\hat{\rho}_{in}$ and the quantity we desire to estimate $\rho_{out}$ which would give us a better picture of the out-of-sample performance of our strategy}.

Maximizing $\hat{\rho}_{in}$ is of course similar to performing a normalized $OLS$, and our interest would be in the analogue of adjustments to MSE. The first is easily found in Mallow's $C_p$. Mallow's formula states that
\begin{eqnarray}
	E[Error^2_{out}] = E[|R-Xb|^2_{out}] \nonumber \\
	= \frac{1}{T}\sum|R_t-X_t b|^2_{in}- Cov(\mu_t,\hat{\mu})_t \nonumber \\
	= \widehat{sse}^2 - \frac{2\sigma_R^2}{N} Tr(M) \nonumber
\end{eqnarray}
where we have confined our attention to linear predictive models, i.e.,
$\hat{mu}=My$
where $M$ is the {\em hat matrix or smooth matrix}.

We recognise that $sse = (1-\rho^2)\sigma_R^2$ from which we can easily derive the following:
\begin{theorem}[Mallow's $C_p$]
	$$E[\rho^2_{out}] = \hat{\rho}^2_{in} - \frac{2}{N} Tr(M)$$ \label{cov_rho}
	where $$R=M X$$ with {\em hat-matrix or smooth} $M$
	%$$E[SR] = \hat{\SR} - \frac{1}{(1+\rho^2)^{3/2}}\frac{2}{N} \tr(M)$$
	
\end{theorem}
We do not prove this since it is a trivial extension of Mallow's. 

If $\hat{\rho}$ is estimated as a by-product of OLS, then $M=X(X'X)^{-1}X$
(see \cite{CASI}, (\S 12.5)) and $Tr(M)=n$, where $k$ is the dimension of the column space of $X$  or the degrees of freedom for regressions (i.e. the total number of lags, where we assume that we have no problems with collinearity). The regularization term turns out to be $\frac{2\cdot k}{N}$ if $\hat{\rho}$ is estimated in-sample by OLS.

If, on the other hand, $\hat{\rho}$ is estimated to optimize the {\sl in-sample} Sharpe ratio, i.e., through TLS, then it is well known that 
$$M=X(X'X-\sigma_{n+1}^2 I)^{-1}X'$$ where $\sigma_{n+1}$ is the smallest singular value of the matrix $T\times (n+1)$ dimensional matrix: $\tilde{X}=[R,X]$ (\cite{Rahman-Yu}). 
As is well known (see for example, \cite{Zhang}) TLS is in fact an anti-regularisation method and may result in less-stable response to outliers. 

In this case, it is also known that $Tr(M)=Tr(Z)$ where $Z= (X'X-\sigma_{n+1}^2 I)^{-1}X'X$ and we know that the singular values of $\sigma(Z)=\{\frac{\lambda_i^2}{(\lambda_i^2-\sigma_{n+1}^2)}\}$ where $\lambda_i$ are the singular values of $X$ (or correspondingly, $\lambda_i^2$ are the singular values of $X'X$, $\lambda_1\geq\cdots\geq\lambda_n>0$ (\cite{Leyang}). 

By the Wilkinson interlacing theorem, $\lambda_n\geq\sigma_{n+1}\geq0$ (see \cite{Rahman-Yu}). Consequently,
$$tr(M)=\sum_i\frac{\lambda_i^2}{(\lambda_i^2-\sigma_{n+1}^2)}>n$$ 
with equality iff $\sigma_{n+1}^2=0$ (i.e., when there the $R^2=100\%$ and consequently, OLS and TLS coincide).  

Thus, while TLS can optimize the {\sl in-sample} Sharpe-ratios over other predictive methods including OLS, the corresponding covariance penalization is larger than that of OLS and {\em may} result in an expected degradation of performance {\sl out-of-sample.}\footnote[4]{For AR models, $k$ is the autoregressive order. For $MA$ models, it is slightly different. For instance, if $M$ is the filter for an EWMA (i.e., an MA(1)), then the $tr(M)$ for the MLE estimate is in fact the normalisation constant. In other words, if $X=(1-\lambda)\sum \lambda^k X_{t-k}$, then $Tr(M)=(1-\lambda)$ rather than the MA order (i.e., 1). This means that a wider  window-size ($\lambda\nearrow 1$) corresponds to fewer effective d.o.f. In the case of the TLS estimated window length, the penalization would be $tr(M)>1$}

The corresponding penalization term can also be applied directly (via first-order conditions or the so-called Delta-method) to removing estimation bias from Sharpe ratio estimates, i.e., 
\begin{eqnarray}
	E[SR_{out}] &\approx& 
	\widehat{SR}_{in} - \frac{2\cdot Tr(M)}{N}\frac{\partial SR_{in}}{\partial\rho}\\
	&\approx& \widehat{SR}_{in} - \frac{2\cdot Tr(M)}{N}\frac{1}{(\hat{\rho}^2+1)^{3/2}}\\
	&\approx &\widehat{SR}_{in} - \frac{2\cdot Tr(M)}{N}(1-\widehat{SR}_{in}^2)^{3/2} \label{cov_imp}
\end{eqnarray}
for all $|\widehat{SR}_{in}|<\frac{\sqrt{2}}{2}$.

We also can extend Stein's Unbiased Regression Estimates (SURE), i.e.
\begin{theorem} SURE
	$$E[\rho^2_{out}] = \hat{\rho}^2_{in} - \frac{2}{N} \sum_{t=1}^N \frac{\partial \hat{X}_t}{\partial R_t} $$
	
\end{theorem}

In both cases, the degrees of freedom are used to adjust down the estimated Sharpe ratios. Consequently, more highly parametrised models will generally have higher overoptimism. The penalisation terms help to adjust for this overoptimism.

\section{Empirical Assessment}

In our empirical investigation, the quantitative strategist is required to build a trading strategy based on lagged information of a certain asset returns. Since the amount of lagged information is not given, he needs to discover it from the given in-sample data. The most naive way (called Naive method in our experiments) is to try different lags and to pick the one that maximizes the Sharpe ratio. Another traditional approach is to select the amount of lags using an information criteria, such as the AIC (Akaike Information Criteria \cite{akaike1974new}) for the OLS case. Nonetheless, we can also apply the Covariance-Penalty methods proposed in the previous section: the Implied Sharpe ratio (Imp SR - eq \ref{cov_imp}) and the R-Squared (eq \ref{cov_rho}) methods.

In summary, our main hypothesis is: 
\begin{itemize}
	\item Can the proposed approaches, Imp SR and R-Squared,outperform AIC and Naive in terms of out-sample/realised Sharpe ratio?
\end{itemize}

In addition to this, we verified the level of alignment between the Expected/In-Sample and Realised/Out-sample Sharpe ratios. With that, our goal was to analyse if there was any additional benefit in terms of performance estimation that can be made by following a Covariance-Penalty approach. In the following subsections we present the data and methodology used to test such hypothesis, as well as the findings and hypothesis testing.

\subsection{Experimental Description}

\subsubsection{Datasets}

Table \ref{datasets} present aggregated statistics associated to the datasets used, whilst Figure \ref{cum_returns_assets_class} present the cumulative returns per asset pool. We have considered three main asset classes during our evaluation: equities, currencies, and fixed income. The data was obtained from Bloomberg, with the full list of 1361 assets tickers and period available at \url{https://www.dropbox.com/s/oogzw8kaysx2qbp/data_list.csv?dl=0}.

\begin{table}[h!]
	\centering
	\tiny
	\caption{Aggregated statistics of the assets used during our empirical evaluation.} \label{datasets}
	\begin{tabular}{p{2cm}|ccccccc}
		\hline
		\hline
		Asset Pool*	& N & Avg Return & Vol & Sharpe Ratio & Calmar Ratio & Monthly Skewness & VaR 95\% \\
		\hline
		All Assets & 1361 & 0.0173 &	0.0425 &	0.4064 &	0.0809 &	-0.9301 &	-0.9314 \\
		World Equity Indices &	18 & 0.0152 & 0.0717 & 0.2114 & 0.0504 & -0.8170 &	-1.0048 \\
		S\&P, SX5E, FTSE, DJIA, Russel and Nikkei Equities & 1289 &	0.0172 &	0.0439 &	0.3915 &	0.0796 &	-0.8816 &	-0.9316 \\
		World Swaps Indices & 14  & -0.0197 & 0.0624 & -0.3149 &	-0.0433 & -0.1039 & -0.9955 \\
		Rates Aggregate Indices & 16 & 0.1220 & 0.0637 & 1.9135 & 0.5504 & -0.8227 &	-0.9440 \\
		World Currencies & 24 & -0.0025 & 0.0315 & -0.0798 & -0.0157 & -1.0052 & -0.8856 \\
		\hline
		\hline
	\end{tabular}
	* Before being averaged, each individual asset was volatility scaled to 10\%
\end{table}

\begin{figure}[h!]
	\centering
	\includegraphics[width=\linewidth]{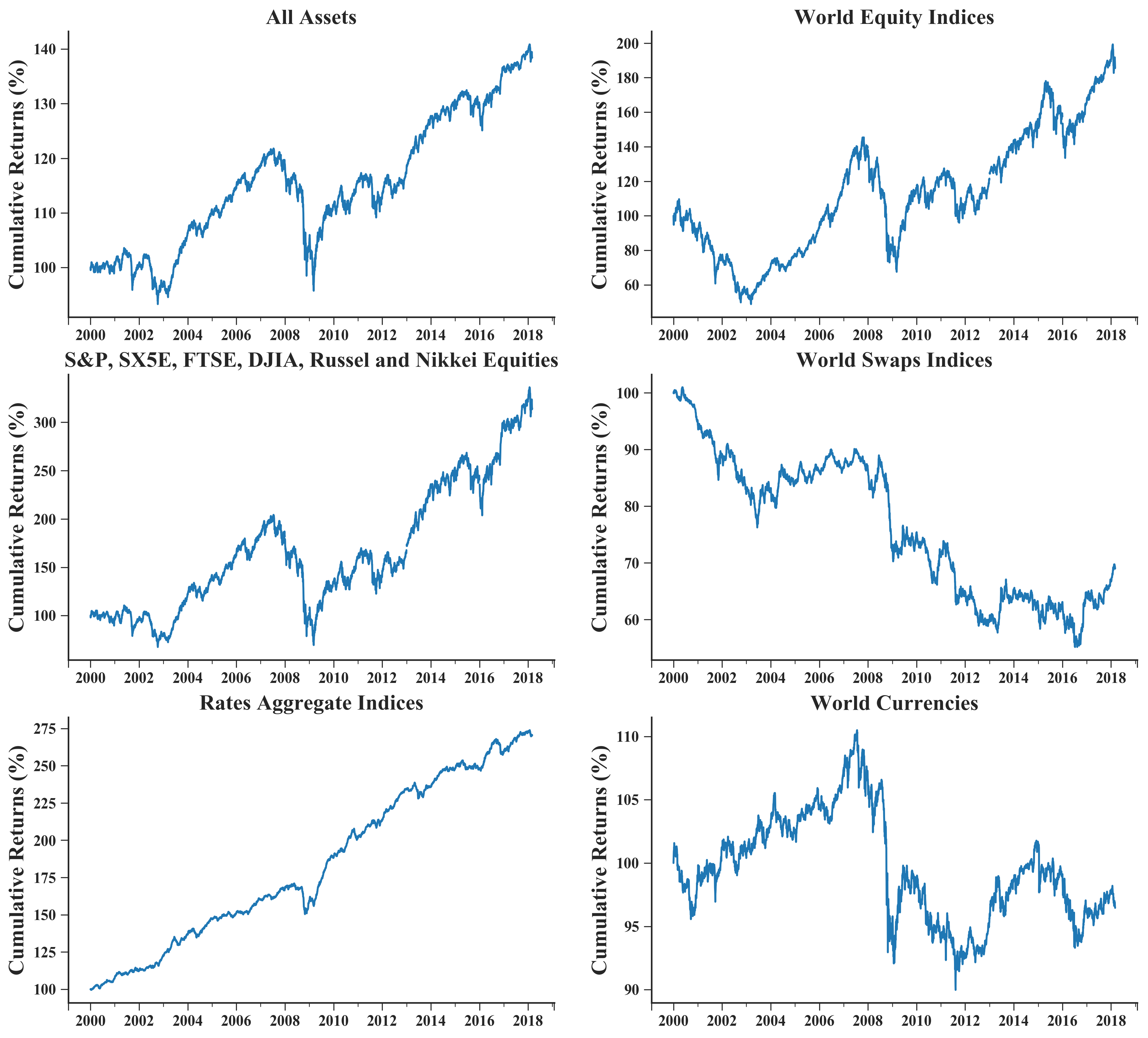} \label{cum_returns_assets_class}
	\caption{Cumulative returns aggregated across asset pool. Before being averaged, each individual asset was volatility scaled to 10\%} \label{cum_returns_assets_class}
\end{figure}

\subsubsection{Model Evaluation Scheme}

\begin{figure}[h!]
	\centering
	\includegraphics[width=\linewidth]{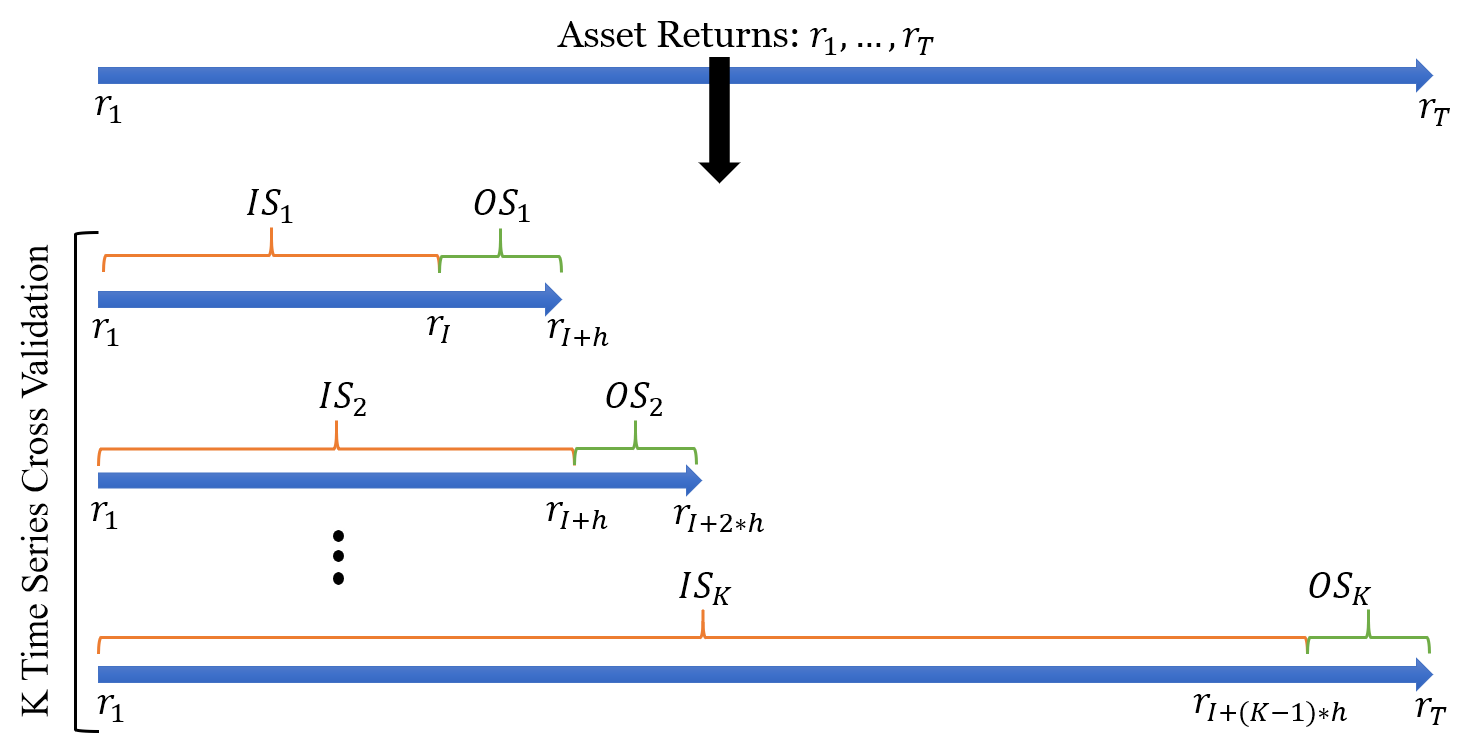}
	\caption{Window-based approach to assess covariance-penalties methodologies.} \label{tscv}
\end{figure}

We have established a testing procedure to assess different Covariance-Penalties methodologies. Figure \ref{tscv} summarize the whole process.

The process start by splitting a sequence of returns $r_1, ..., r_T$ in several in-sample ($IS$) and out-sample ($OS$) sets. The first pair, $IS_1$ and $OS_1$, are determined by the trading horizon $h$ and the first $IS$ batch size $I$. After that, the whole sequence of $IS_k$ and $OS_k$ are created by simply accumulating $IS_k = \{IS_{k-1}, OS_{k-1}\}$ and rolling $OS_k = \{r_{I + (k-1)*h}, ..., r_{I + k*h}\}$ until $I + k*h \leq T$. Every Covariance-Penalty methodology -- Naive, Implied SR, R-squared and AIC -- tap into the $IS$ data to fine-tune a particular trading strategy. The measure of their success to avoid overfit is obtained when this strategy is applied in the $OS$ set. Using the sequence of in-sample and out-sample results we are able to compute performance metrics to evaluate and test the different hypothesis. We have used $h=21$ days and the initial period $I=1008$ days in our experiments; this process is repeated per dataset and Covariance-Penalty.

\subsubsection{Trading Strategy and Performance Metrics}

Below we present the steps taken to backtest the different trading strategies. For every pair of in-sample and out-sample data $(IS_k, OS_k)$, run:

\begin{itemize}
	
	\item For every lag $p \in \{3, 5, 7, 9, 12, 15, 18, 21, 26, 31, 36, 42, 49, 56, 63, 84, 105, 126\}$, do:
	
	\begin{enumerate}
		\item Fit the linear model using the in-sample set available ($IS_k$):
		\begin{equation}
		r_t = \beta_0 + \beta_1 r_{t-1} + \beta_2 r_{t-2} ... + \beta_p r_{t-p} + \varepsilon_t = \hat{r}_t + \varepsilon_t \nonumber
		\end{equation}
		with the coefficients $\beta_0, \beta_1, ...$ identified via Ordinary or Total Least Squares.
		
		\item Compute the risk of choosing $p$ using the in-sample set:
		
		\begin{equation}
		L_p = \mathcal{L}(..., (\hat{r}_{t-1}, r_{t-1}), (\hat{r}_{t}, r_{t}), (\hat{r}_{t+1}, r_{t+1}), ...; M) \nonumber
		\end{equation}
		with hat matrix $M$ and $P$ as a performance function, representing the different types of penalization used (AIC, Naive, Imp SR and R-squared). For the AIC and R-Squared it was also computed the expected Sharpe ratio ($\hat{SR}_p$).
		
	\end{enumerate}
	
	\item Pick $p^* = argmax_p \{L_3, L_5, ..., L_{126}\}$, multiplying by $-1$ in the AIC case.
	
	\item Fit the linear model using $p^*$ and the in-sample set available ($IS_k$):
	\begin{equation}
	r_t = \beta_0 + \beta_1 r_{t-1} + \beta_2 r_{t-2} ... + \beta_{p^*} r_{t-{p^*}} + \varepsilon_t = \hat{r}_t + \varepsilon_t \nonumber
	\end{equation}
	
	\item Test this model against the out-sample data ($OS_k$). Store out-sample realised strategy returns, $L_{p*}^{(k)}$ and $\hat{SR}_{p*}^{(k)}$.
	
\end{itemize}

With the sequence of results per asset in batches of in and out-sample, $\{(\hat{r}_{I}, r_I)$, $(\hat{r}_{I+1}, r_{I+1}), ...,$ $(\hat{r}_{I+h*k}, r_{I+h*k}); L_{p*}^{(k)}; \hat{SR}_{p*}^{(k)}\}_{k=1}^{K-1}$, we computed some metrics related to the alignment between what was Expected and Realised:

\begin{itemize}
	\item Correlation ($\rho$ as the Pearson correlation coefficient):
	\begin{equation}
	Correlation = (\frac{1}{K-1}) \sum_{i=1}^{K-1} \rho(SR^{(1:i)}, \hat{SR}_{p*}^{(k)}) 
	\end{equation}
	\item Mean Absolute Difference ($MAD$)
	\begin{equation}
	MAD = (\frac{1}{K-1}) \sum_{i=1}^{K-1} |SR^{(1:i)} - \hat{SR}_{p*}^{(k)}| 
	\end{equation}
\end{itemize}

where $SR^{(1:i)}$ is the Out-sample/Realised Sharpe ratio of a trading strategy until out-sample set $i$. Hence, the Correlation measures the association between the Expected and Realised Sharpe ratios, whilst $MAD$ is measuring its distance; the best case being $MAD = 0$ and $Correlation = 1$.

Finally, using the sequence as a whole per asset, $\{(\hat{r}_{I}, r_I)$, $(\hat{r}_{I+1}, r_{I+1})$, $..., (\hat{r}_{I+h*(K-1)}, $ $r_{I+h*(K-1)})\}$, we computed the Realised Sharpe Ratio. In all computations made, the Sharpe ratio was computed considering a 3-month Libor rate as the benchmark rate. 

\subsection{Results and Discussions}

\begin{figure}[h!]
	\centering
	\includegraphics[width=\linewidth]{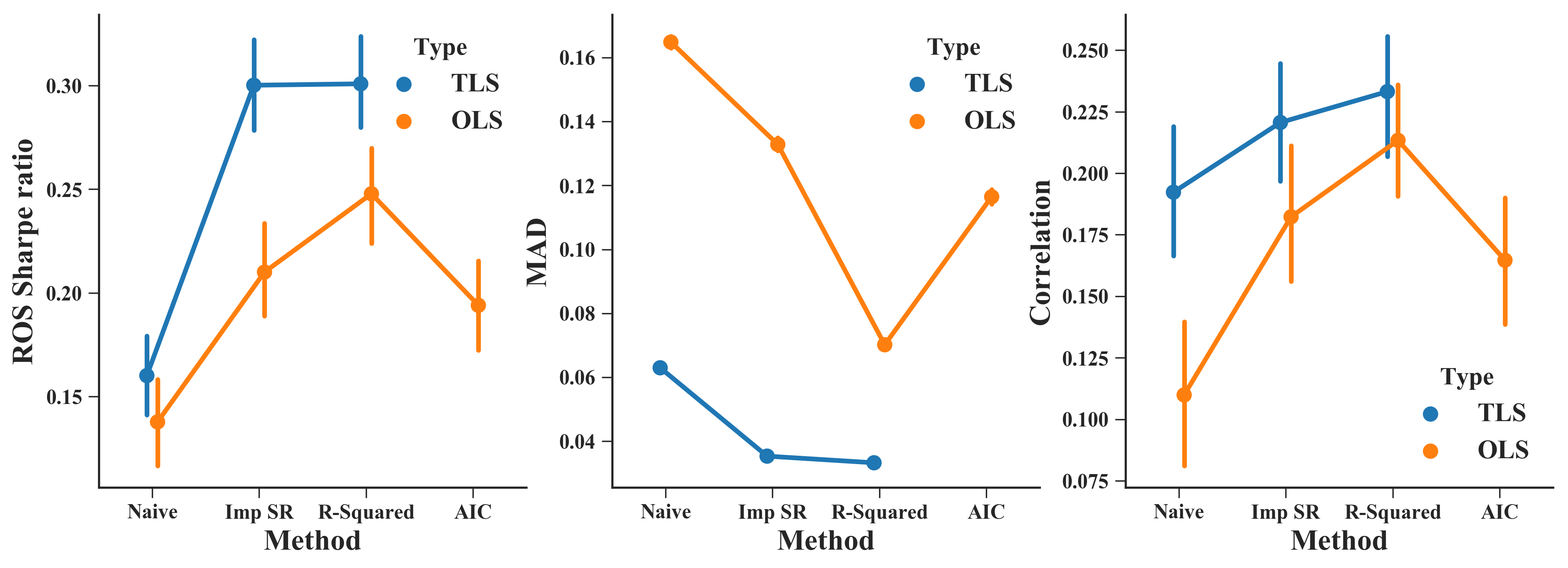}
	\caption{Aggregated results for the different Covariance-Penalty methods, and types of Least Squares: Realised Out-Sample (ROS) Sharpe ratio, MAD and Correlation. The dot and bar represent the mean and standard error, respectively, across 1361 assets.} \label{covpen_perf_ossr_corr_mad}
\end{figure}

Figure \ref{covpen_perf_ossr_corr_mad} present the aggregated results for the different Covariance-Penalty methods, and types of Least Squares across the 1361 assets. In terms of Realised Out-Sample (ROS) Sharpe ratio, using some Covariance-Penalty method provides a boost in the unconditional average of 80-100\% percent in relation to the Naive approach. Our two proposed approaches, Imp SR and R-squared, also improves 20-30\% in relation to the AIC lag selection. Overall, this pattern is also visible in terms of alignment between Expected and Realised Sharpe ratios: MAD is reduced and Correlation is increased. 

To check the main hypothesis, we have performed a statistical comparison between all approaches in terms of their ROS Sharpe ratio. Table \ref{statscomp} present the results of paired t-tests, controlling for multiple hypothesis testing using Bonferroni correction. Overall, three main patterns can be spotted:

\begin{itemize}
	\item All Covariance-Penalty approaches outperformed significantly the Naive approach; and our proposed two, Imp SR and R-Squared, also compares favourably in relation to the AIC.
	\item R-Squared provided slightly better results than Imp SR, but its effect-size is small when compared to AIC and Naive cases.
	\item TLS improves ROS Sharpe ratio in relation to OLS; as we have outlined before, since Sharpe ratio and Correlation are strongly associated, as TLS is maximizing Correlation, it will tend to perform better than OLS in relation to this metric. 
\end{itemize}

\begin{table}[h!]
	\footnotesize
	\centering
	\caption{Statistical comparisons between different Covariance-Penalty approaches and the baseline (Naive), using paired t-test and Bonferroni Correction for multiple hypothesis test. The boldfaced pairs are the ones not significantly different at 5\% after Bonferroni correction.} \label{statscomp}
	\begin{tabular}{cc|cc|cc}
		\hline
		\hline
		\multicolumn{2}{c}{Val Method} & \multicolumn{2}{c}{ROS Sharpe ratio} & \multicolumn{2}{c}{p-value} \\
		\hline
		Left & Right & Mean diff & t-stat & Value & Bonferroni  \\
		\hline
		OLS-AIC &       OLS-Imp SR &                -0.0235 &               -3.7573 &  0.00018 &             0.00376 \\
		OLS-AIC  &      OLS-Naive   &                0.0575   &               10.061  & $<$ 0.00001   &      $<$ 0.00001 \\
		OLS-AIC   & OLS-R-Squared    &             -0.05351    &            -8.6793 & $<$ 0.00001     &    $<$ 0.00001 \\
		OLS-AIC    &   TLS-Imp SR     &             -0.1036     &           -12.0062 & $<$ 0.00001 &       $<$ 0.00001 \\
		OLS-AIC     &   TLS-Naive      &            0.03583      &           3.7806 & 0.000163  &     0.003430 \\
		OLS-AIC &   TLS-R-Squared       &           -0.1048       &         -12.1588 & $<$ 0.00001   &   $<$ 0.00001 \\
		OLS-Imp SR  &      OLS-Naive        &          0.08118        &         13.8264 & $<$ 0.00001    &       $<$ 0.00001 \\
		OLS-Imp SR  &  OLS-R-Squared         &         -0.0299         &       -4.3723 & 0.00002 &         0.00028 \\
		OLS-Imp SR   &    TLS-Imp SR          &        -0.0800          &      -9.4425 & $<$ 0.00001  &       $<$ 0.00001 \\
		OLS-Imp SR    &    TLS-Naive           &       0.05942           &      6.2843 & $<$ 0.00001   &     $<$ 0.00001 \\
		OLS-Imp SR &   TLS-R-Squared             &    -0.08122             &   -9.5304 & $<$ 0.00001     &  $<$ 0.00001 \\
		OLS-Naive  &  OLS-R-Squared              &    -0.1111              &  -14.7347 & $<$ 0.00001 &    $<$ 0.00001 \\
		OLS-Naive   &    TLS-Imp SR               &   -0.1612               & -17.9283 & $<$ 0.00001  &         $<$ 0.00001 \\
		\textbf{OLS-Naive}    &    \textbf{TLS-Naive}    & -0.02176    &-2.4326  &  0.01512   &   \textbf{0.31753} \\
		OLS-Naive &   TLS-R-Squared                 & -0.1624     &           -17.9653 & $<$ 0.00001    &    $<$ 0.00001 \\
		OLS-R-Squared  &     TLS-Imp SR                  &-0.0501      &          -6.3812 & $<$ 0.00001 &       $<$ 0.00001 \\
		OLS-R-Squared   &     TLS-Naive   &               0.08934       &          9.0067 & $<$ 0.00001  &    $<$ 0.00001 \\
		OLS-R-Squared    & TLS-R-Squared   &              -0.05129       &         -6.5487 & $<$ 0.00001  &  $<$ 0.00001 \\
		TLS-Imp SR      &  TLS-Naive     &              0.1394         &        14.8044 & $<$ 0.00001    &  $<$ 0.00001 \\
		\textbf{TLS-Imp SR} &   \textbf{TLS-R-Squared}      &     -0.00118  &     -0.4581  &   0.64694 &  \textbf{1.00000} \\
		TLS-Naive  &  TLS-R-Squared       &           -0.1406           &     -14.7307 & $<$ 0.00001  &  $<$ 0.00001 \\
		\hline
		\hline
	\end{tabular}
\end{table}

\begin{figure}[h!]
	\centering
	\includegraphics[width=\linewidth]{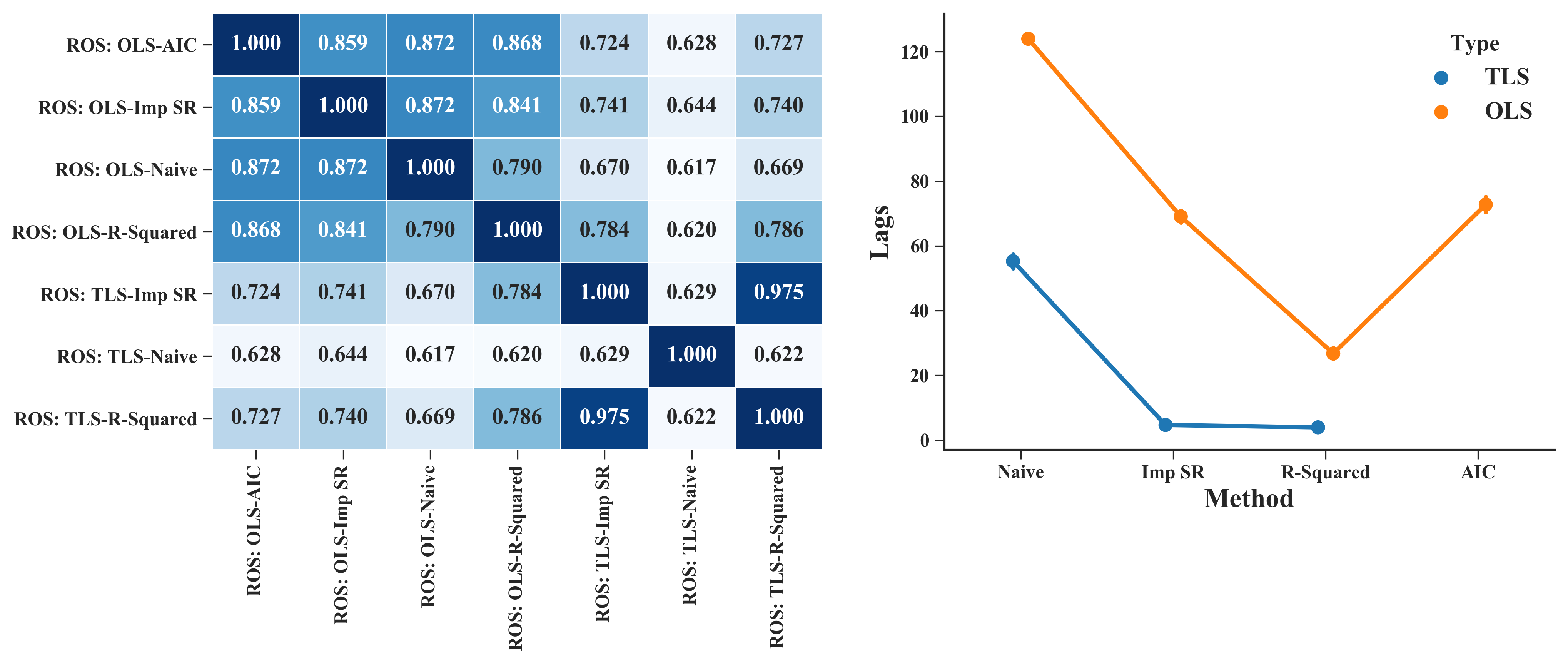} \label{covpen_roscorr_hyperparams}
	\caption{Correlation matrix between ROS Sharpe ratio across methods and types, with the average most frequent (mode) lag selected across 1361 assets.}
\end{figure}

There is also another explanation of why TLS has outperformed OLS: Figure \ref{covpen_roscorr_hyperparams} outline the similarity between approaches in terms of ROS Sharpe ration, and the most frequent (mode) lag selected across the in-out sample sets, averaged across 1361 assets. Overall, TLS penalized harder than OLS, making it many setups to adopt smaller lags, which in some sense, mimic a mean-reversion strategies; similarly, whenever OLS preferred smaller Hypothesis Spaces (less parameters), ROS Sharpe ratio rose. Overall ROS Sharpe ratio are quite correlated across the different methods and types, but some diversification might still  be attained by pooling TLS and OLS models. 

\newpage
\section{Conclusions}

In this paper we proposed a new approach to dealing with financial overfitting, a Covariance-Penalty Correction, in which a risk metric is lowered given the amount of parameters and data used to underpins a trading strategy. We outlined the foundation and main results behind the Covariance-Penalty correction for trading strategies. After that, we pursued an empirical investigation, comparing its performance with some other approaches in the realm of Covariance-Penalties across more than 1300 assets, using Ordinary and Total Least Squares -- with the last providing superior performance. Before delving into this novel perspective, we first briefly exhibit a historical context about the available methodologies. 

We outlined, chronologically, three distinct approaches in the literature to evaluate and deal with backtesting overfitting: Data Snooping, Overestimated Performance, and Cross-Validation Evaluation. The Data Snooping and Overestimated Performance authors identify that the root cause of any overfitting or spurious results is due to the multiple trials (hypothesis) employed by the analyst during the strategies fine-tuning step. In contrast, the Cross-Validation Evaluation and Covariance-Penalty Correction are focused on avoiding the apparent in-sample performance, by uncovering the expected generalisation (out-sample) potential of a strategy. The Covariance-Penalty Correction methods require some assumptions on the joint behaviour of the asset and trading strategy, which is not the case for the majority of the Cross-Validation procedures. However, this lack of assumptions demand an extra computational processing in respect to the Covariance-Penalty approach.

We then focused on to address the issue of overfitting of dynamic models, we look to estimated objective functions, taking degrees of freedom into account. Most commonly used estimates include Akaike and Bayes' information criteria, but these require the use of a likelihood. In our case, we do not have an exact likelihood, but instead choose to optimise the Sharpe ratio, and noting the monotonicity in $\rho$, this comes from maximizing $\hat{\rho}$. Consequently, we wish to consider the means of optimising correlation taking into account penalisations for degrees of freedom. In this sense, we presented formulas for adjusting estimated in-sample goodness of fits measures, to produce better estimates of the out-of-sample Sharpe ratio performance. Effectively, we would like to have a more sober estimate of the potential performance of our optimised strategies out of sample.

We then performed an empirical investigation, in which the quantitative strategist is required to build a trading strategy based on lagged information of a certain asset returns. The most naive way (called Naive method in our experiments) is to try different lags and to pick the one that maximizes the Sharpe ratio. We compared it with the Akaike Information Criteria (AIC), as well as with the Covariance-Penalty methods proposed in this work: the Implied Sharpe ratio (Imp SR) and the R-Squared methods. In summary, our main hypothesis is: Can the proposed approaches, Imp SR and R-Squared,outperform AIC and Naive in terms of out-sample/realised Sharpe ratio? In addition to this, we verified the level of alignment between the Expected/In-Sample and Realised/Out-sample Sharpe ratios. With that, our goal was to analyse if there was any additional benefit in terms of performance estimation that can be made by following a Covariance-Penalty approach. 

Our findings suggest that Realised Out-Sample (ROS) Sharpe ratio, using some Covariance-Penalty method provides a boost in the unconditional average of 80-100\% percent in relation to the Naive approach. Our two proposed approaches, Imp SR and R-squared, also improves 20-30\% in relation to the AIC lag selection. Overall, this pattern is also visible in terms of alignment between Expected and Realised Sharpe ratios: MAD is reduced and Correlation is increased. Overall, all Covariance-Penalty approaches outperformed significantly the Naive approach; and our proposed two, Imp SR and R-Squared, also compares favourably in relation to the AIC. Total Least Squares (TLS) improves ROS Sharpe ratio in relation to Ordinary Least Squares (OLS). One of the explanations why TLS outcompeted OLS was that TLS penalized harder than OLS, making it many setups to adopt smaller lags, which in some sense, mimic a mean-reversion strategies. Similarly, whenever OLS preferred smaller Hypothesis Spaces (less parameters), ROS Sharpe ratio rose. Overall ROS Sharpe ratio are quite correlated across the different methods and types, but some diversification might still  be attained by pooling TLS and OLS models. 

\section*{Acknowledgement}

Adriano Koshiyama would like to acknowledge the National Research Council of Brazil for his PhD scholarship.

\bibliographystyle{plain}
\bibliography{bibthesis}

\appendix

\section{Full distributions for single period
} \label{Section:FullDist}

\begin{figure}[h]
	\centering
	\includegraphics[width=0.6\linewidth]{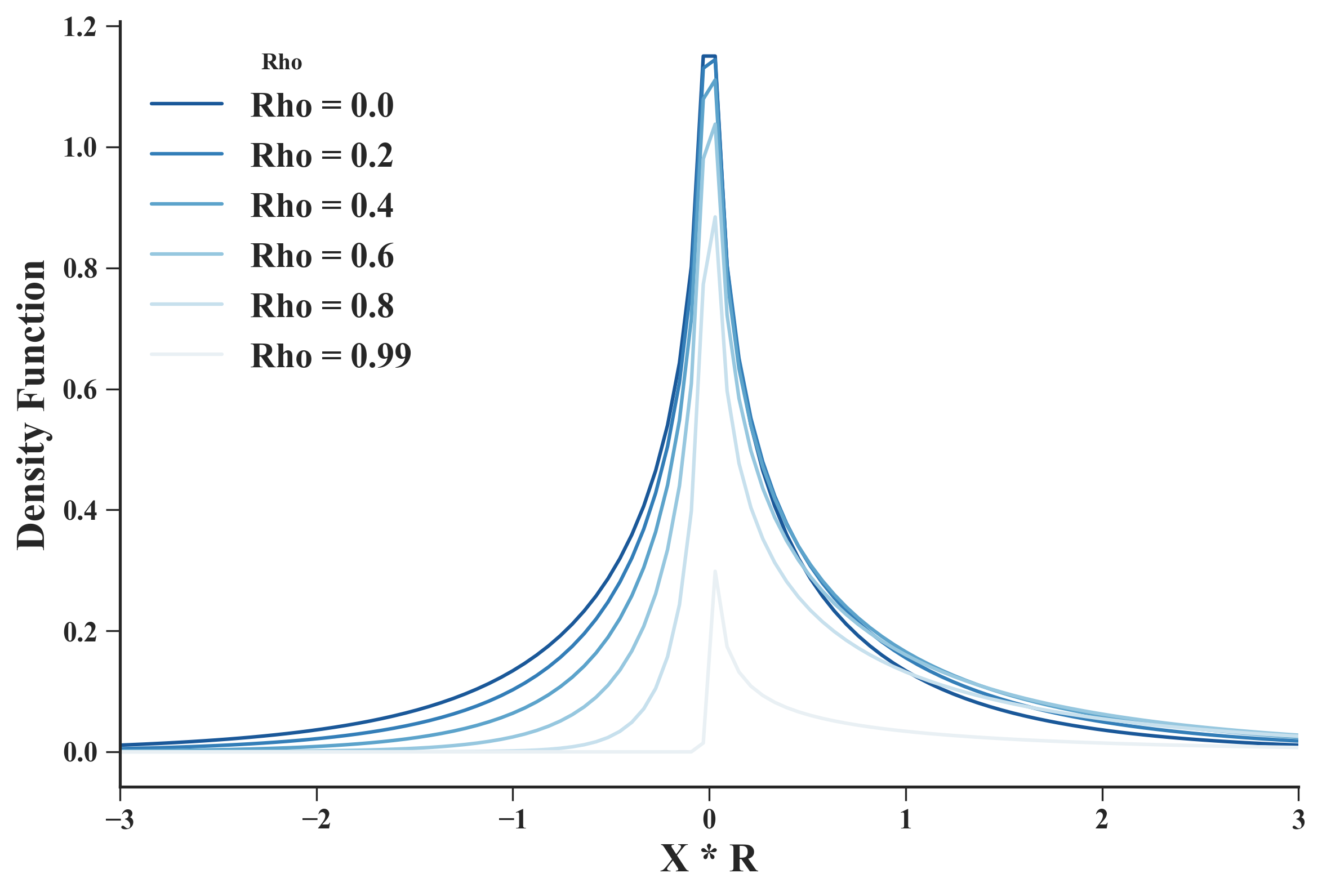}
	\caption{Complete product distributions for $\rho\in\{0,0.2,0.4,0.6.,0.8,1.0\}$, normalised to have unit variance
	}
	\label{fig:exact}
\end{figure}

In general, for $X$ and $R$ jointly normal, we
have $X_{t}R_{t}$ is known to have a pdf, 

\[
p_{x}=\frac{1}{\pi\sigma_R\sigma_X}\exp\bigl(\frac{\rho x}{\sigma_R\sigma_X(1-\rho^{2})}\bigr)K_{0}\bigl(\frac{|x|}{\sigma_R\sigma_X(1-\rho^{2})}\bigr)
\]
where $K_0(\cdot)$ is a modified Bessel function of the $2^{nd}$ kind  (\cite{Simons}, p 51, eq 6.15). The more general density for non-zero means, is given in \cite{Exact} as an infinite series. In the special cases of independence and of correlated but zero mean, the expressions become much simpler and we choose to focus on the zero-mean case here. The density is unbounded at zero and has fat tails and positive skewness, becoming more pronounced with higher correlation. We can see the distribution for a variety of correlations in figure (\ref{fig:exact}), with the skewness becoming increasingly pronounced for higher $\rho$. In the limit as $\rho\rightarrow 1$ the distribution converges to that of the central $\chi^2$ distribution with one degree of freedom.

\section{ Nonzero means: Sharpe ratios and Skewness}

By an abuse of notation, we define $SR[R]$ to be $\mu_R/\sigma_R$ and by an abuse of notation, we define $SR[X]=\mu_X/\sigma_X$ (for $X$ the signal),

{\bf Corollary 1:} If $R\sim \mathscr{N}(\mu_R,\sigma_R^{2})$ and $X\sim \mathscr{N}(\mu_X,\sigma_X^{2})$ then
$$ SR[X\cdot R] = {SR[R]\cdot SR[X] + \rho \over (SR[R]^2+SR[X]^2+2\rho SR[R]\cdot SR[S]+\rho^2+1)^{1/2}}
$$

{\bf Corollary 2:} If $R\sim \mathscr{N}(\mu_R,\sigma_R^{2})$ and $X\sim \mathscr{N}(\mu_X,\sigma_X^{2})$  then
$$ \gamma_3[X\cdot R] = { 2\rho (\rho^2 + 3 + 3 SR[R]^2+3 SR[X]^2)\over
	(SR[R]^2+SR[X]^2+2\rho SR[R]\cdot SR[X]+\rho^2+1)^{3/2}}
$$

We note  the one period Sharpe ratio of the 	strategy may depend on both the interaction between the Sharpe ratios of the Signals (weights) and the Returns, in particular whether they have the same sign or not, together with the sign of the correlation. In fact, the amplitude of the resulting strategy SR may be more dependent on the respective Sharpe ratios rather than $\rho$ since after all, $-1\leq\rho\leq 1,$while $SR[R]$ and $SR[X]$ may individually be above $1$.

\end{document}